\documentclass[10pt,a4paper]{article}
\usepackage{amsmath}
\usepackage{amscd}
\usepackage{amssymb}
\usepackage{graphicx}
\usepackage{latexsym}
\usepackage{color}
\usepackage{floatrow}
\usepackage{colortbl}
\usepackage{authblk}
\usepackage[left=3cm,right=3cm,top=3cm,bottom=3cm]{geometry}

\title{Extra-cellular matrix rigidity may dictate the fate of injury outcome\\}

\author[1]{D.~Peurichard}
\author[2]{M. Ousset}
\author[2]{J. Paupert}
\author[3]{B.~Aymard}
\author[2]{A. Lorsignol}
\author[2]{L.~Casteilla}
\author[4]{P.~Degond}

\affil[1]{\small INRIA Paris, 2, rue Simone Iff, 75589 Paris Cedex 12, France}
\affil[2]{\small STROMALab, Université de Toulouse, Inserm U1031, EFS, INP-ENVT,  UPS, CNRS  ERL5311, Toulouse, France Batiment INCERE, 4 bis Avenue Hubert Curien 31100 Toulouse, 31 432 Toulouse– France}
\affil[3]{\small MathNeuro Team, Inria Sophia Antipolis Méditerranée, 2004 Route des Lucioles, BP93, 06902 Valbonne cedex, France}
\affil[4]{\small Department of Mathematics, Imperial College London, London SW7 2AZ, United Kingdom.}
\date{}

\begin{document}
\maketitle

\begin{abstract}
After injury, while regeneration can be observed in hydra, planaria and some vertebrates, regeneration is rare in mammals and particularly in humans. In this paper, we investigate the mechanisms by which biological tissues recover after injury. We explore this question on adipose tissue, using the mathematical framework recently developed in Peurichard et al., J. Theoret. Biol. 429 (2017), pp. 61-81. Our assumption is that simple mechanical cues between the Extra-Cellular Matrix (ECM) and differentiated cells can explain adipose tissue morphogenesis and that regeneration requires after injury the same mechanisms. We validate this hypothesis by means of a two-dimensional Individual Based Model (IBM) of interacting adipocytes and ECM fiber elements. The model successfully generates regeneration or scar formation as functions of few key parameters, and seems to indicate that the fate of injury outcome could be mainly due to ECM rigidity.\\

\end{abstract}

\section{Introduction}

The discoveries of stem cells largely open the door for the development of regenerative medicine and address many issues on regeneration processes and their basic cues. Regeneration is described as the come back to the initial shape and function of a tissue or organ after injury. Largely described and investigated in animal models such as hydra, zebra fish or salamander, regeneration was considered as largely impaired in adult mammals for the benefit of a fibrotic scar after healing obligatory associated with dysfunctions. In mammals, regenerative capacities seem present until a few days after birth and then further replaced by healing and scar formation in adult. This strongly suggests that regeneration could take place in adult mammals but is inhibited during postnatal development \cite{Aymard2016,Ailhaud_1999,Alonso_2014} and that fibrosis impairs tissue regeneration. The possibility to circumvent this inhibition and to identify basic cues to re-induce regeneration in adults represents a fundamental challenge in regenerative medicine. One of these issues is to investigate whether basic cues driving the development of mature tissues and organs could be similarly involved in the post-injury rebuilding of these mature tissues. Because of the severe pandemic rise of obesity and several associated diseases (diabetes, cardiovascular diseases, metabolic disorders, cancer), adipose tissue (AT) has become a growing point of interest in the last decades. Previously considered as a passive energy store, AT is now recognized to be an important endocrine organ involved in all physiological functions \cite{Kershaw_2004}. Beside this historical role related to its metabolic functions, adipose pad is classically removed in obese people or injected in patients for plastic and reconstructive surgery although, its true regenerative capabilities after lipectomy are poorly documented \cite{Seretis_2015}. It also recently emerges as a reservoir of therapeutic multipotent adipose-derived stromal/stem cells (ASCs) for regenerative medicine \cite{Nambu2009,Vindigni2012}. These cells have been extensively investigated and increasingly tested and used in clinical trials \cite{Monsarrat_2016}. Therefore, understanding the mechanisms involved in adipose tissue morphogenesis and homeostasis is of major importance either to limit its expansion or to promote and regenerate it according to the applicative fields.  Alternatively and independently of these applied interests, AT can be considered as a model because it is a relatively simple organ, composed of mature adipocytes with a spherical shape composed of lipid droplets storing excess energy and a population of other cells (endothelial, immune cells, fibroblast-like cells [57]) among which ASCs that behave as adipocyte progenitors. In white AT, the differentiated adipocytes are densely packed together in specific structures referred to as lobules, surrounded by an organized (aligned) collagen fiber network \cite{Vincent_1991}. These cellular and fibrous structures show different organization according to the location in the tissue as well as according to states of development and pathologies. These global organization and structure largely impact the functioning of the tissue and metabolic dys-regulations of AT are associated with the development of fibrosis in obese patients \cite{Friedl_2000,Friedl_2000,Pellegrinelli_2014}. 

Assuming that the structuring agents contributing the most to adipose tissue emergence are the Extra-Cellular Matrix (ECM) fibers and the adipocytes interactions \cite{Peurichard2016}, we proposed a 2D individual based model (IBM) which produces structures quantitatively similar { (shape of the cell clusters in particular) } to the experimental observations of healthy AT in adults. This model indicates that AT homeostasis and organization could spontaneously emerge as a result of simple mechanical interactions between differentiated cells (i.e. adipocytes) and fibers. { Our model bears analogies with the model of \cite{Buske2012}, in which an individual based model for the simulation of intestinal organoid formation in the gut is discussed. The main differences between our model and that of \cite{Buske2012} lie in the way the Extra-Cellular Matrix and its interactions with the cells are modelled. In \cite{Buske2012}, the basal membrane is described by a stiff triangulated network modelling interlinked long polymer molecules. The network edges can stretch or compress and the network nodes may move in response to the forces exerted by the cells. However, network reorganization does not occur spontaneously but only when some faces are too large or too small, which triggers the insertion or deletion of a vertex. The present model allows for ECM spontaneous reorganization via the creation and suppression of links dynamically in time. Thus, in the present paper, the ECM has the ability to reorganize independently from the action of the cells. This is a key feature which allows us to explore how the ECM remodelling capability influences the regeneration process. Another difference is cell adhesion to the ECM which is key to \cite{Buske2012} and is not considered here, because cell-fiber repulsion is sufficient to provide containment of cells.   }

\begin{figure}
\includegraphics[scale = 	0.25]{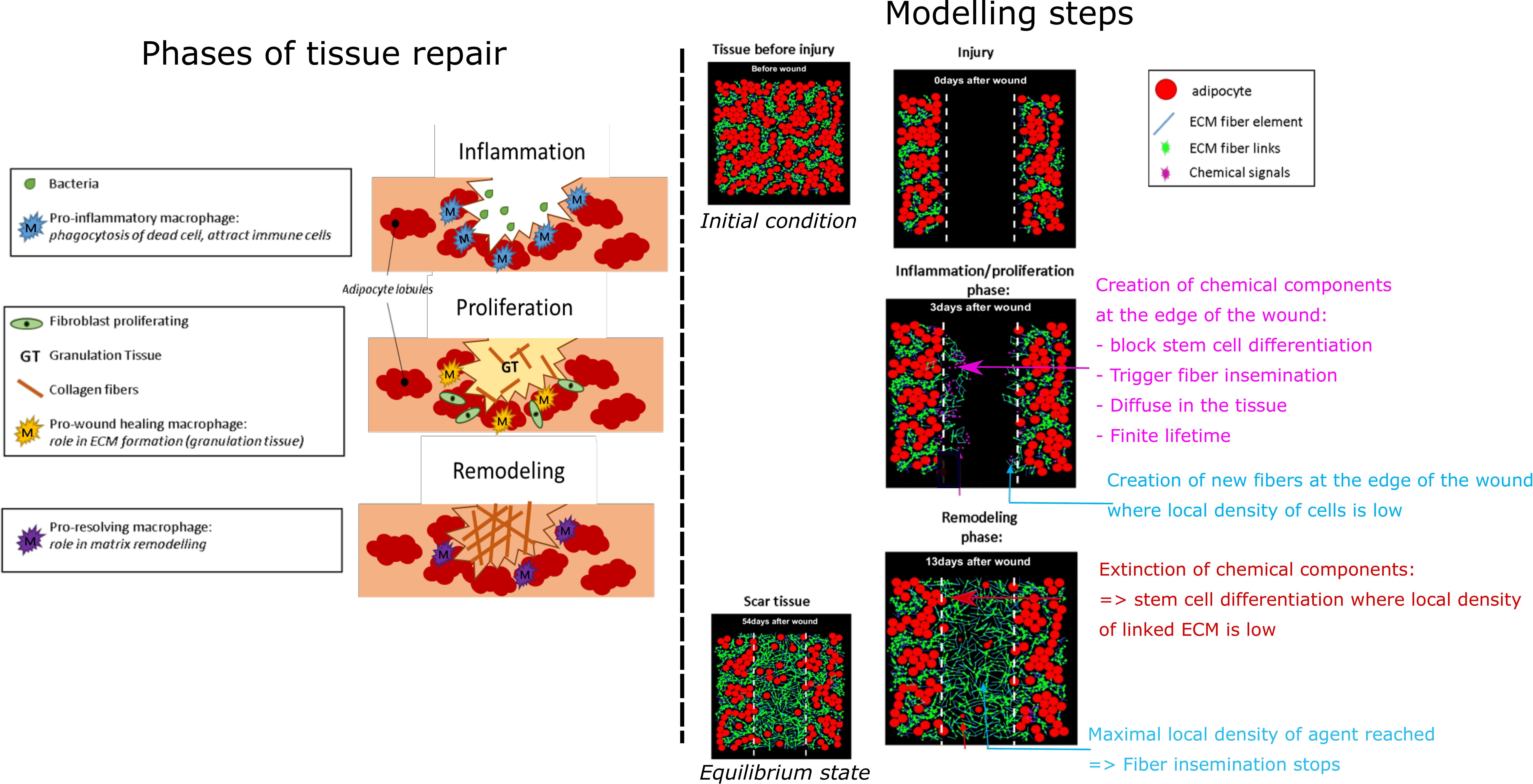}
\caption{Left: Schematic view of the three phases of tissue repair: Hemostasis/inflammation, proliferation and remodeling/maturation. Right: Schematic view of the model components at different time points.\label{Fig1scheme}}
\end{figure}

The goal of the present study is, via a mathematical model based on our previously developed 2D-IBM and using the same framework, to investigate whether the same rules drive the fate of AT reconstruction after injury. To this aim, we first summarize the main steps of tissue repair, which consist of three phases { (summarized in Fig. \ref{Fig1scheme} left)} : Hemostasis/inflammation, proliferation and remodeling/maturation. The first stage of acute tissue repair is dedicated to stop bleeding (hemostasis) and damaged-cell death. Cell death is characterized by rupture of the cell membrane and release of intracellular factors. This release induces an acute inflammation which is associated to specific immune cell type recruitment responsible for phagocytosis of dead cell debris but also for the production of the anti-inflammatory cytokines required for the down-regulation of the inflammatory response and necessary to prevent further inflammation-related damages. Progressive resolution of inflammation is ultimately responsible for the passage from a necrotic environment to one favorable to stem cells homing, proliferation and differentiation necessary to tissue repair. This proliferation stage is associated to granulation tissue formation characterized by deposition of an immature provisional ECM, composed of poorly organized fibrous components. In parallel, this environment induces fibroblasts differentiation into myofibroblasts, responsible of fibrous collagen synthesis, which marks the transition to remodeling stage. This maturation phase allows terminal differentiation of progenitor cells and remodeling of collagen-enriched ECM. ECM remodeling and maturation in a strongly organized network is controlled by its synthesis, its degradation properties as well as by its post-translational modifications such as fiber cross-linking ensured by specific enzymes. This remodeling is a key step in driving tissue regeneration since collagen accumulation upon tissue repair is associated with fibrotic scar formation. This fibrosis alters tissue mechanical properties especially increasing its stiffness and eventually impairs tissue regeneration.

The model developed in this paper is based on an existing model \cite{Peurichard2016} which we modify to take into consideration the previously described phases of tissue repair. In section 2., we will first present the main ingredients of the model and motivate the choice of the model parameter values with biological measurements (section 2). Section 3. is devoted to numerical results: in sections 3.1 and 3.2, we show numerical results which we compare qualitatively to biological observations, and in section 3.3 we quantify the numerical structures obtained to study the influence of key model parameters on the tissue reconstruction abilities after injury.
 
\section{Methods}

\subsection{Biological experiments}

\begin{figure}[h]
\includegraphics[scale = 0.3]{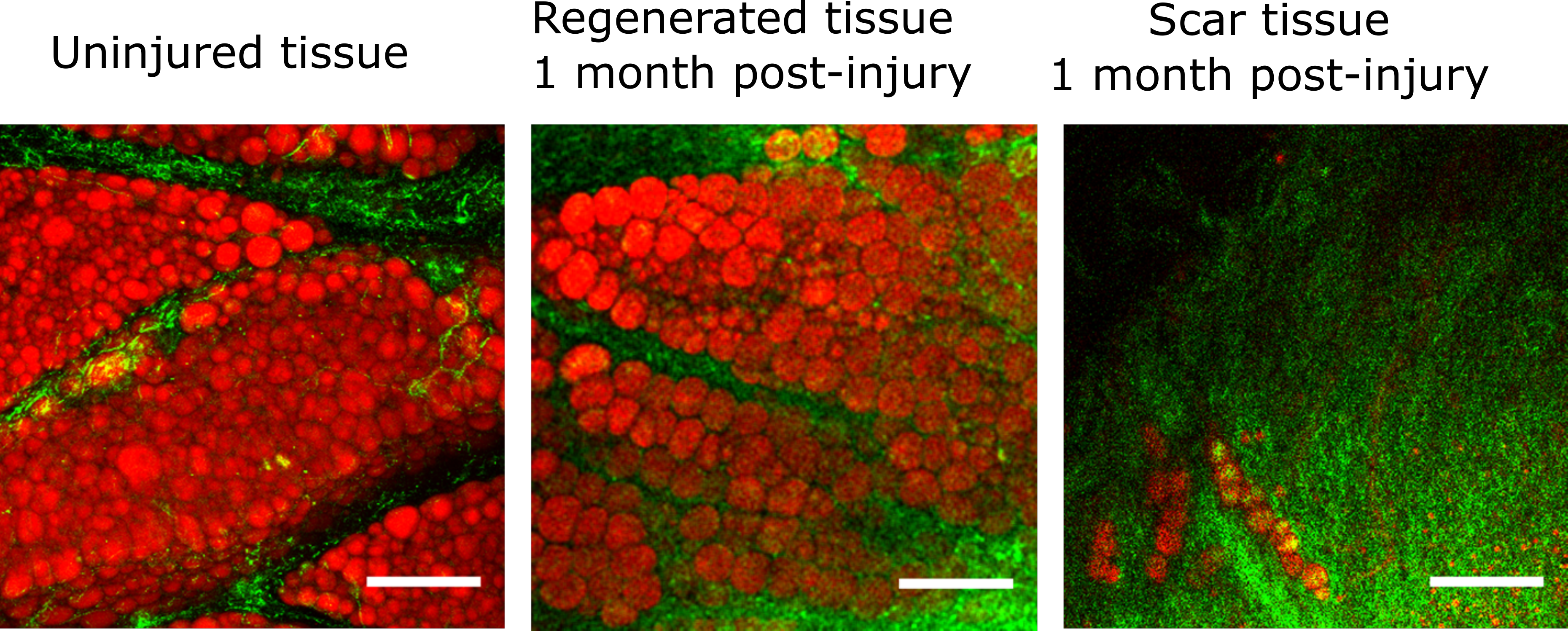}
\caption{ Biological images of inguinal adipose tissue (AT) in non-injured conditions (A) and in regenerative (B) or scar (C) conditions one month after injury. Adipocytes (Red) were stained with Bodipy and collagen fibers (green) were imaged using collagen SHG (Second Harmonic Generation) signal. Images were obtained using average intensity projections of 8 (A), 7(B) and 5(C) stack images. The white scale bars at the bottom right correspond to $150 \mu m$. \label{Fig1}}
\end{figure}

{ In Fig. \ref{Fig1}, we show biological images of inguinal adipose tissue in non-injured conditions (A) and in regenerative (B) or scar (C) conditions one month after injury. Adipocytes (Red) were stained with Bodipy and collagen fibers (green) were imaged using collagen SHG (Second Harmonic Generation) signal (see Appendix \ref{method} for more details on the methods). As one can observe in Figs. \ref{Fig1}, the adipocyte lobules and ECM structures are reconstructed 1 month post-injury for regenerative tissues while the injured zone in scar conditions is filled with a very stiff ECM network with very few adipocytes.  }
\subsection{Description of the model}

(i) Introduction of new fibers/cells. { To create a simulation framework that addresses the mechanical features of adipogenesis and homeostasis, a 2D individual based model (IBM) was introduced in \cite{Peurichard2016}. Here, we use the same framework to model tissue repair after injury. ECM fibers are discretized into unit fiber elements consisting of line segments of fixed and uniform lengths represented by their centers and their directional unit vectors. In the previous model, we supposed that two fiber elements that cross each other can form links, thereby creating longer fibers. By analogy with what is observed in vivo, these segments represent collagen fibers as they are secreted in the extracellular compartment in vivo and their ability to be cross-linked. The cells/adipocytes are described as 2D spheres represented by their centers and radii. The differentiation of immature cells into adipocytes is represented by their ability to grow regularly in size (hypertrophy). At any given time, the two sets of agents are supposed to realize the minimum of the mechanical energy of the system (further described). To model tissue injury, we start with situations of tissues at equilibrium, and remove all the components (cells and fibers) present in a given zone of the domain. Thus, to drive the emergence of fibers at the leading edge of the wound, we create a gradient which triggers fiber insemination. This gradient is generated by a flux of chemical components produced by surrounding cells, mimicking secretions of cells acting during inflammatory and proliferation phases { (see Fig. \ref{Fig1scheme} right)}. The wound is filled with new fibers and the injury signaling chemicals disappear, enabling maturation of new adipocytes. All these phenomena force the cells and fibers to move in order to reach a new equilibrium. New fibers are inseminated randomly in time via a Poisson process of frequency $\nu_f(X)$ which negatively depends on the agent local density and its gradient and which has typical value $\nu_f^*$ , an agent being either a cell or a fiber element. Therefore, fiber insemination is supposed to occur more likely on the boundaries of the wound and where the agent density does not exceed a given threshold (to avoid accumulation of fibers). { As we do not model proliferating fibroblasts for the sake of simplicity, the collagen fiber elements are supposed to spontaneously emerge at the borders of the wound}. Fiber elements in our model are not intended to reproduce a whole biological structured fiber too long to be represented by a single short fiber element but they rather produce a discretization of such fibers into rigid elements that can be computationally manipulated. At initialization, the insemination of a new fiber is modeled by the insemination of a group of 4 fiber elements organized in linked 'diamond' configurations corresponding to non-orientated collagen fiber secretion. The major axis orientation of the diamonds is randomly chosen in $[ 0; 2\pi ]$ with uniform probability. Moreover, we suppose that newly inseminated fibers can link with their neighbors when they intersect with a fixed probability $P_f$, at the time they are created. Note that these links are dynamical in time, due to the linking/unlinking process (further described). Therefore, the diamond shape structures might be dissolved or persist according to how fast the unlinking process is. Considering an initial fiber network consisting of fibers longer than cell diameters is biologically plausible as collagen fibers are produced before any implantation of adipocytes.

Simultaneously with fiber insemination, we suppose that the agents at the leading edge of the wound create a flux of injury signaling chemicals. These signals are modeled as particles created with the same rules as newly inseminated fibers (described before), with space-dependent insemination frequency $\nu_R(X)$ having typical value $\nu_R^*$. They are supposed to have a finite lifetime: injury signaling particles die according to a Poisson process of finite frequency $\nu_D$. These injury-signaling particles are supposed to diffuse through the ECM in a directed manner, the direction being given by the mean local orientation of the fibrous tissue { (comprised between $-\pi/2$ and $\pi/2$ as the fiber elements are not directed)}. { The dynamics of these signaling particles are described by a continuum model (shown in the Appendix), discretized using particles following a Smoothed-Particle Hydrodynamics (SPH) approach. The SPH method performs better than classical grid methods in the presence of zero-density regions where traditional grid-based methods are likely to generate unphysical negative densities which severly compromise the simulation. With SPH methods, zero-density regions are simply devoid of particles. Here, the density of injury signalling molecules is initially zero, and there remain large zero-density regions in the simulation domain throughout the computation, which makes the choice of the SPH method a necessity.  } Following [43], we suppose that the stem cells have a negligible impact on the mechanical equilibrium and we do not incorporate them in the model. The differentiation of a stem cell into an adipocyte is modeled as the creation (or 'insemination') of a new adipocyte, of space-dependent frequency $\nu_a(X)$ having typical value $\nu_a^*$. All new adipocytes are inseminated with the same small radius. In order to delay cell differentiation during tissue reconstruction, new adipocytes are inseminated at random times following a Poisson process the frequency of which depends on: (a) The density of injury-signaling particles: New adipocytes appear more likely where the density of injury signaling particle is low, (b) the number of fiber links in the ECM: new adipocytes appear more likely in regions of low ECM fiber links and (c) the density of existing agents: new adipocytes appear more likely where existing fibers are already present. The differentiation yield is modeled through the regular growth of the cells. We assume that the volume of each adipocyte reaches a maximal value beyond which it stays constant. Because adipocytes are filled with an incompressible liquid we assume that two neighboring disks can not overlap. 

(ii) Fiber-adipocyte interactions: We view the ECM as composed of fibers having locally preferred directions. Each fiber element exerts an elementary mechanical action corresponding to a fraction of the action the ECM exerts on neighboring cells. It carries directional information corresponding to the preferred direction of the collagen fibers. The largest mechanical action of a fiber is exerted normally to this direction. The fibers repel the cells by means of a soft potential allowing some penetration of the cells inside the ECM. A fiber element produces a unit of potential strength. Larger mechanical actions are achieved by having several of these unit fiber elements in a close neighborhood. In practice, we assume that the fiber-cell repulsion potential isolines are ellipses with foci at the two ends of the fiber segment and that the potential vanishes beyond a certain distance from the fiber. The choice of elliptic isolines corresponds to the simplest anisotropic shape in 2D. 

(iii) Fiber growth, elongation and ability to bend: In addition to carrying a unit of ECM fiber strength, fiber elements also carry a unit of fiber length. However, fiber elongation [8] is created by allowing two crossing fibers to create a cross-link. Any displacement of a cross-linked fiber pair maintains the position of the cross-link relative to the center of each fiber. Several consecutively cross-linked fiber elements model a long fiber having the ability to bend or even take possible tortuous geometries. Pairs of cross-linked fibers can also spontaneously unlink, allowing for fiber breakage describing ECM remodeling processes. Linking and unlinking processes follow Poisson processes with frequencies $\nu_\ell$ and $\nu_d$ respectively, and the parameter $\chi_\ell = \frac{\nu_\ell}{\nu_\ell + \nu_d}$ , where $\chi \in [0, \; 1]$ represents a measure of the fraction of linked fibers among the pairs of intersecting fibers. To model the ability of long fibers (those made of several cross-linked fiber units) to offer a certain resistance to bending, linked fibers are subjected to a potential torque at their junction. This torque vanishes when the fibers are aligned, and consequently acts as a linked fibers alignment mechanism. Any force exerted by a cell in the vicinity of a cross-link would result in the cross-linked fibers making an obtuse angle with respect to one another until the exerted torque at the cross-link balances the effect of the force, thereby providing a discrete representation of the bending of a continuous beam. This torque is characterized by a parameter $c_1 > 0$ playing the role of a flexural modulus. This parameter reflects the linking forces as well as the number of links existing between fibers. The larger the $c_1$ the less elastic the fiber assembly is. Finally, steric interactions among the fibers are considered by adding a fiber-fiber mechanical repulsion. Each fiber element represents the elementary mechanical action acting on neighboring fibers. The fibers repel each other by means of a soft potential carrying directional information and the potential is stronger in the orthogonal direction of the fibers.

The $N_A$ adipocytes are modeled as 2D spheres of center $X_i$ and time dependent radius $R_i$ for $i \in [1,\; N_A]$, and the $N_f$ ECM fiber elements are represented by straight lines of fixed length $L_f$, of center $Y_k$ and orientation angle $\theta_k$ for $k \in [1,\; N_f ]$. We generate an injury that occurs at a chosen simulation time (in which the tissue is at mechanical equilibrium) and leads to the complete destruction of half of the whole tissue. We delete all the elements (cells and fibers) the centers of which are contained in $[- \frac{x_{max}}{2}, \; \frac{x_{max}}{2} ] \times [-y_{max}, \; y_{max} ]$ where the 2D square simulation domain $\Omega = [- x_{max}, \; x_{max} ] \times [-y_{max}, \; y_{max} ]$. The details of the model are presented in Appendix C.

\subsection{Biological relevance of the model parameters}
The reference time $t_{\text{ref}}$ is chosen to be 0.05 times the cell growth time $t_g$, i.e the time for which a cell reaches its maximal radius. As it takes approximately 100 days for a nascent adipocyte to grow to its maximal size and $t_g = 20$ in our simulations, we can assume that $t_{\text{ref}}$ is about 5 days \cite{Arner_2010}. Simulations are stopped at dimensionless time $t = 80t_{\text{ref}}$. In the following, we consider three fiber unlinking times $t_d = \frac{1}{\nu_d} $: $t_d = \{10, 100, 1000\}t_{\text{ref}}$, and the linking time $t_\ell = \frac{\chi_\ell}{1 - \chi_\ell} t_d$ (we use $\chi_\ell = 0.35$ or 0.1). Note that these are linking/unlinking times per fiber but the actual frequencies
of linking/unlinking events must be multiplied by the number of fibers elements $N_f$. As in \cite{Peurichard2016}, we can estimate that there is significant ECM remodelling when at least $10\%$ of the fibers (i.e about 100 fibers) have been linked or unlinked. Therefore, the ECM remodelling rate can be related to $100 \nu_d$ and consequently, the ECM remodelling time $t_d/100$ ranges between $0.1$ and $10$ days. In \cite{Cowin_2004}, it is estimated that ECM remodelling takes about 15 days. Therefore we will refer to the first cases as 'highly dynamical tissues' and the last ones as 'regular tissues'. The choice of fiber length and width is motivated by studies of \cite{Ushiki_2002}, where it is shown that the collagen fibers are organized into bundles of collagen fibrils the width of which varying from $1$ to $20 \mu m$ depending on tissues and organs. In order to enable the fibers to bend around the cells, the fiber elements are chosen to be $20 \mu m$ wide and $60 \mu m$ long (the adipocyte diameter being 60$\mu m$, see \cite{Peurichard2016}). 
The model parameters used for the simulations are given in table \ref{Table}.

\section{Results}
 
The simulations are performed on a 2D square domain with periodic boundary conditions. Each simulation starts from a tissue at equilibrium that we cut in half to mimic an injury. We consider two types of tissues: (i) 'Regular tissues' with low remodelling rate, i.e with a low cross-linked fiber unlinking frequency $\nu_d = 10^{-3}$ and (ii) 'Highly dynamical tissues' with $\nu_d = 10^{-1}$. Note that the ECM remodelling parameters $\chi_\ell,\nu_d$ and $\nu_\ell$ are the same for computing the tissue homeostasis phase before injury and for the reconstruction phase after injury. To obtain the tissue before injury, we start the simulations with a fiber network randomly distributed and randomly linked at $40 \%$ (see table \ref{Table}) and let the simulation run until reaching equilibrium (see \cite{Peurichard2016}). The cell insemination frequency parameter $\nu_a^*$, the injury signaling chemicals insemination rate $\nu_R^*$ and death rate $\nu_D^*$ and the sensitivity to the matrix local number of links for cell insemination are fixed, and we study the influence of the fiber insemination frequency parameter $\nu_f^*$ and of the linking probability of new fibers $\mathbf{P}_f$ on the structures obtained after injury, { for two different types of tissues (playing on the dynamics of the tissue via the unlinking frequency $\nu_d$)}.

\subsection{Regular tissues}
In Fig.\ref{Fig:chi01nud1e31} (I) (see also \ref{S1_Vid}), we show simulation results for $\chi_\ell = 0.1$ and $\nu_d = 10^{-3}$ at different times $t=0.2 t_{\text{ref}}$, $t=0.4t_{\text{ref}}$, $t=10t_{\text{ref}}$, $t=20t_{\text{ref}}$ and $t=80t_{\text{ref}}$, for low fiber insemination rate after injury $\nu_f^*=0.25$ and moderate linking of new fibers $\mathbf{P}_f = 0.2$. 
\begin{figure}[h!]
\includegraphics[scale=0.3]{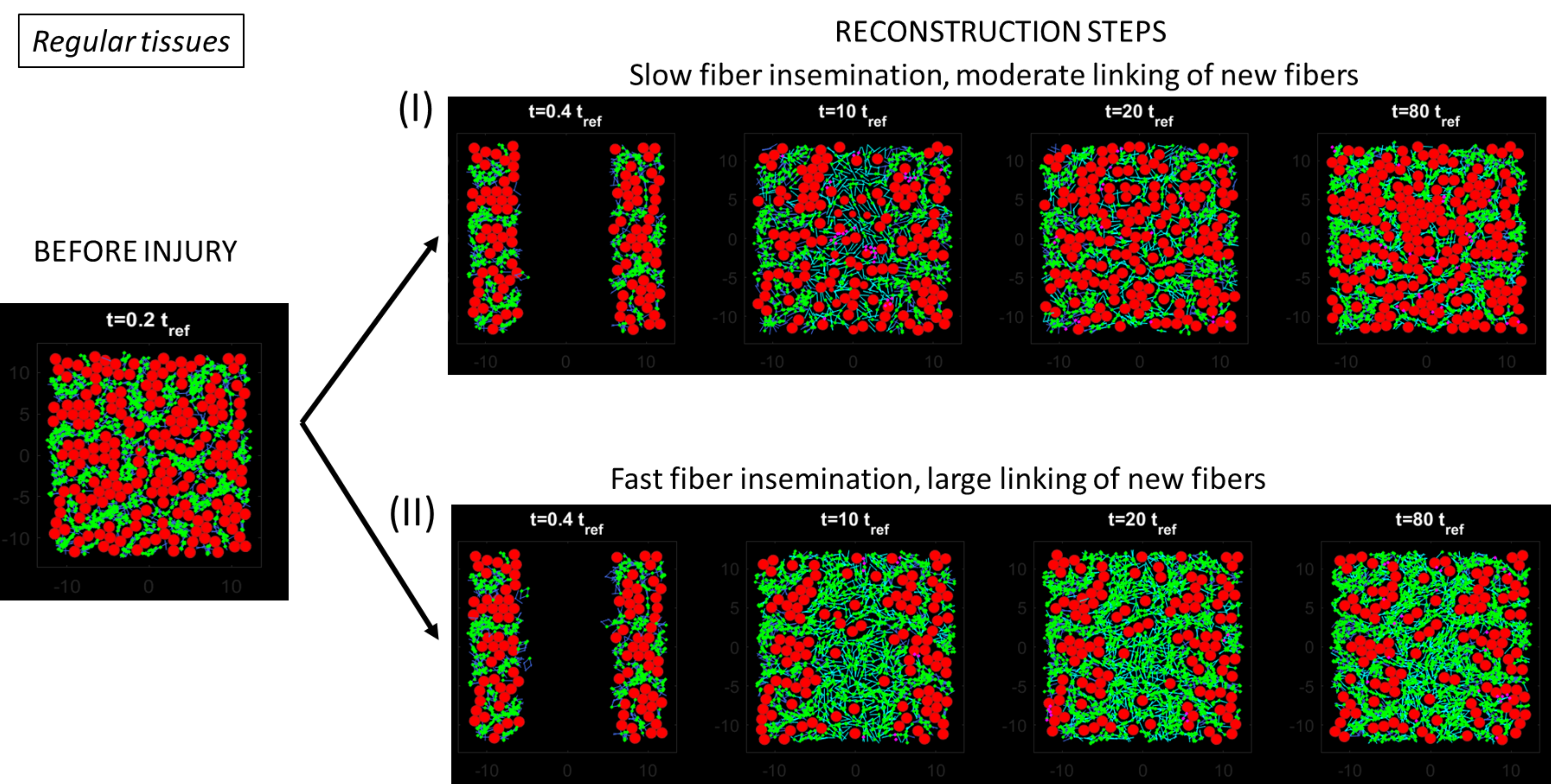}
\caption{Left panel: Simulation results before injury for tissue with low fiber remodelling rate: $\chi_\ell = 0.1$, $\nu_d = 10^{-3}$. Right panel: Reconstruction steps (I) We consider slow fiber reconstruction rate $\nu_f^* = 0.25$ and moderate linking of new fibers $\mathbf{P}_f=0.2$, (II) We consider fast fiber reconstruction rate $\nu_f^* = 2$ and large linking of new fibers $\mathbf{P}_f=0.5$. From left to right: at time $t=0.2t_{\text{ref}}$ (before injury), $t=0.4t_{\text{ref}}$, $t=10t_{\text{ref}}$, $t=20t_{\text{ref}}$ and $t=80 t_{\text{ref}}$. \label{Fig:chi01nud1e31}}
\end{figure}
As shown in Fig. \ref{Fig:chi01nud1e31}, for slow fiber creation and moderate fiber linking, lobule-like clusters of cells emerge in the reconstructed zone, with similar features as before the injury and same number of adipocytes as before injury (compare left and right images of Fig. \ref{Fig:chi01nud1e31}). 

{ In Fig. \ref{Fig:chi01nud1e31} (II) (see also \ref{S2_Vid}), we use the same initial condition, i.e with simulation parameters $\chi_\ell=0.1$, $\nu_d = 10^{-3}$, but with different ECM reconstruction parameters: We consider a fast fiber insemination with rate $\nu_f^* = 2$ with large linking probability of new fibers $\mathbf{P}_f = 0.5$. In this case, the injured zone is filled with a highly interconnected fiber network possessing very few adipocytes, structure which resembles a scar. These results show that the fiber network reconstruction speed and reconstructed ECM can trigger regeneration or scar formation in 'regular networks'. In the next section, we study the case of highly dynamical networks.}

\textit{Remark: Note that simulations of this section are performed for tissues with low remodelling rate $\nu_d~=10^{-3}$, i.e each fiber link has a large lifetime. For such slow remodelling network, the total number of links is not proportional to $\chi_\ell$ but rather close to the initial linking proportion, in this case around $40\%$ of crossed-fibers. To compare this tissue with more dynamical ones, we will use the value $\chi_\ell = 0.35$ for tissues with larger $\nu_d$ (see next section).}
 
\subsection{Highly dynamical tissues}
In this section, we consider what is referred to as 'highly dynamical networks', i.e fiber networks with fast remodelling rate $\nu_d = 10^{-1}$. In order to compare these tissues with the ones considered in the previous section, we use the value $\chi_\ell = 0.35$ such that the two compared tissues have the same proportion of links before injury (around $35 \%$ of crossed-fibers are linked before injury).

In Fig. \ref{Fig:chi035nud1e11} (I) (see also \ref{S3_Vid}), we show simulations results for slow fiber insemination rate after injury $\nu_f^* = 0.25$ and moderate linking probability of newly inseminated fibers $\mathbf{P}_f = 0.2$. After injury, the tissue is regenerated in the sense that we observe the same structures as before injury for this set of parameters. 
\begin{figure}[h!]
\includegraphics[scale=0.3]{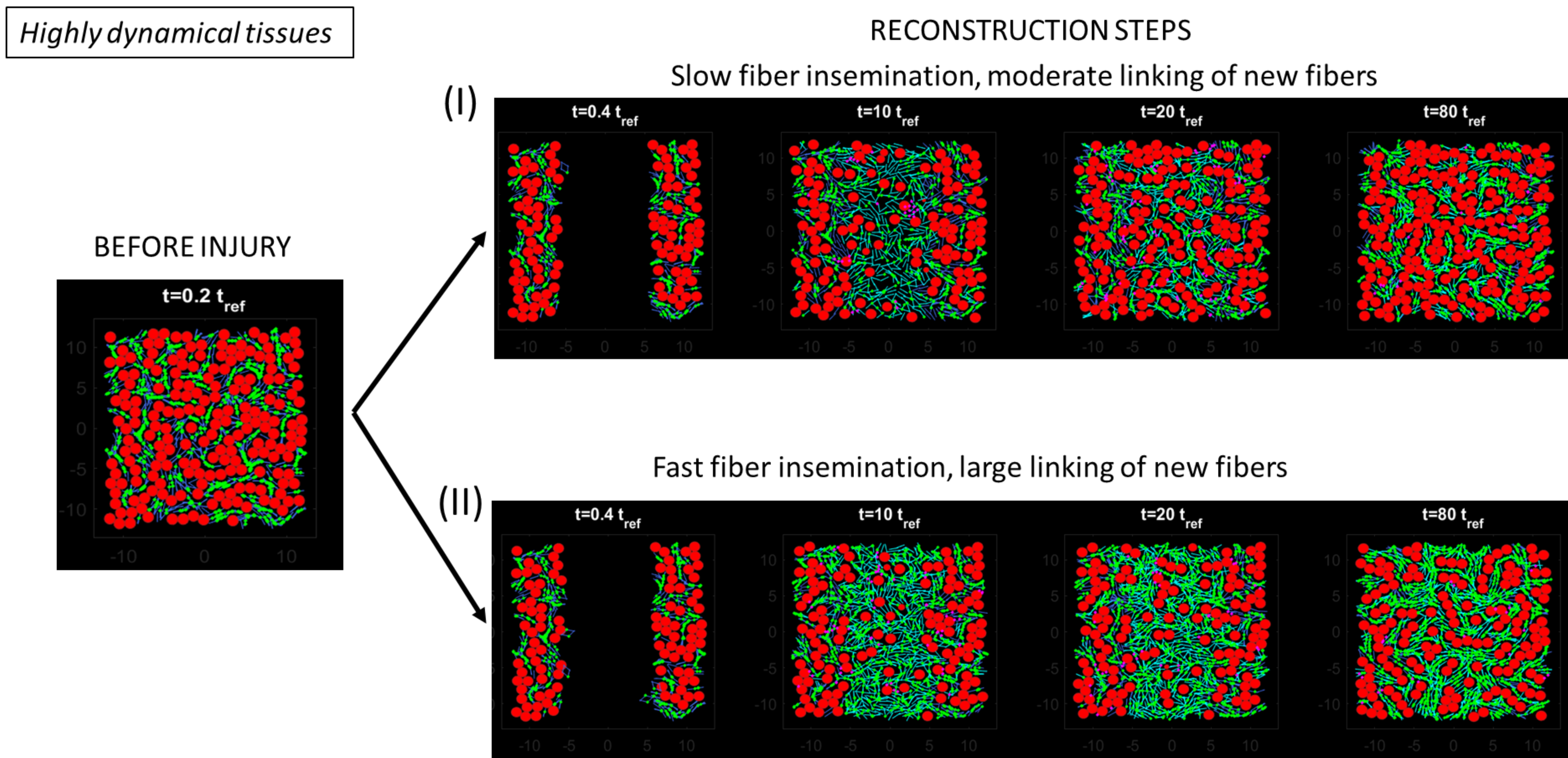}
\caption{Left panel: Simulation results before injury for tissue with low fiber remodelling rate: $\chi_\ell = 0.35$, $\nu_d = 1$. Right panel: Reconstruction steps (I) We consider slow fiber reconstruction rate $\nu_f^* = 0.25$ and moderate linking of new fibers $\mathbf{P}_f=0.2$, (II) We consider fast fiber reconstruction rate $\nu_f^* = 2$ and large linking of new fibers $\mathbf{P}_f=0.5$. From left to right: at time $t=0.2t_{\text{ref}}$ (before injury), $t=0.4t_{\text{ref}}$, $t=10t_{\text{ref}}$, $t=20t_{\text{ref}}$ and $t=80 t_{\text{ref}}$.\label{Fig:chi035nud1e11}}
\end{figure}

Finally, coupling fast ECM reconstruction with large linking probability of newly inseminated fibers ($\nu_f^* = 2$, $\mathbf{P}_f = 0.5$) -shown in Fig.\ref{Fig:chi035nud1e11} (II) (see also \ref{S4_Vid})- leads to the emergence of a highly interconnected and organized (aligned) fiber network containing few adipocytes. 

When comparing the two reconstructed tissues displayed in Fig.\ref{Fig:chi035nud1e11} (II right) and in Fig. \ref{Fig:chi01nud1e31} (II, right), one can note that highly dynamical networks (Fig. \ref{Fig:chi035nud1e11} II) do not enable the formation of a scar even when considering fast creation of highly interconnected fibers, contrary to the observations in 'regular tissues' (Fig.\ref{Fig:chi01nud1e31} II). This is due to the fact that highly dynamical tissues rapidly remodel the fiber network, thereby creating organized (aligned) fiber structures. By rapidly creating and suppressing new links, these networks have a larger plasticity than the 'regular tissues'. As stem cell differentiation is sensitive to the local number of links in the ECM network, this results in the creation of new adipocytes in these tissues.

\subsection{Parametric analysis of the model}
In order to enable the quantitative comparison between the tissue states before and after injury, we proceed as in \cite{Peurichard2016} and we define quantifiers $E$ and $N_C$ measuring the elongation and number of cell clusters respectively, and $A$ measuring the mean alignment of fiber clusters (see \ref{computationSQ} for more details on their computation). As we are interested in comparing the reconstructed tissue with the state before injury (to identify cases of regeneration or scar formation), we define the 'global error' ${\Upsilon}$ as a weighted norm of the quantifiers before and after injury:
$$
\Upsilon = \sqrt{\big(\frac{\Delta E}{E_0}\big)^2+\big(\frac{\Delta A}{A_0}\big)^2+\big(\frac{\Delta N_C}{N_{C_0}}\big)^2+\big(\frac{\Delta N_E}{N_{E_0}}\big)^2},
$$
\noindent where $N_E$ is the number of cells and $\Delta X = X_{80t_{\text{ref}}} - X_{0}$ for all quantifiers $X$. We choose $E_0=0.6$, $A_0=0.2$, $N_{C_0}=10$ and $N_{E_0} = 0.2$ for all simulations, which correspond to the values of the difference of the quantifiers between a regular tissue and a scar. Thus defined, large values of $\Upsilon$ indicate that the reconstructed tissue is far from its initial state (before injury), while small values of $\Upsilon$ indicate that the final structures resemble the ones before injury. As functions of the ECM reconstruction parameters, we observe three different final states: (i) Regeneration, i.e complete reconstruction of the tissue after injury with the same number of adipocytes and formation of lobule-like clusters of cells in the new network (low values of $\Upsilon$), (ii) formation of an unstructured tissue with the same number of cells as before injury, but which does not display any organization of the cell and fiber clusters and (iii) scar formation, where the injured zone is filled with highly interconnected and disorganized fibers with very few adipocytes (large values of $\Upsilon$). 
\begin{figure}[h!]
\includegraphics[scale=0.4]{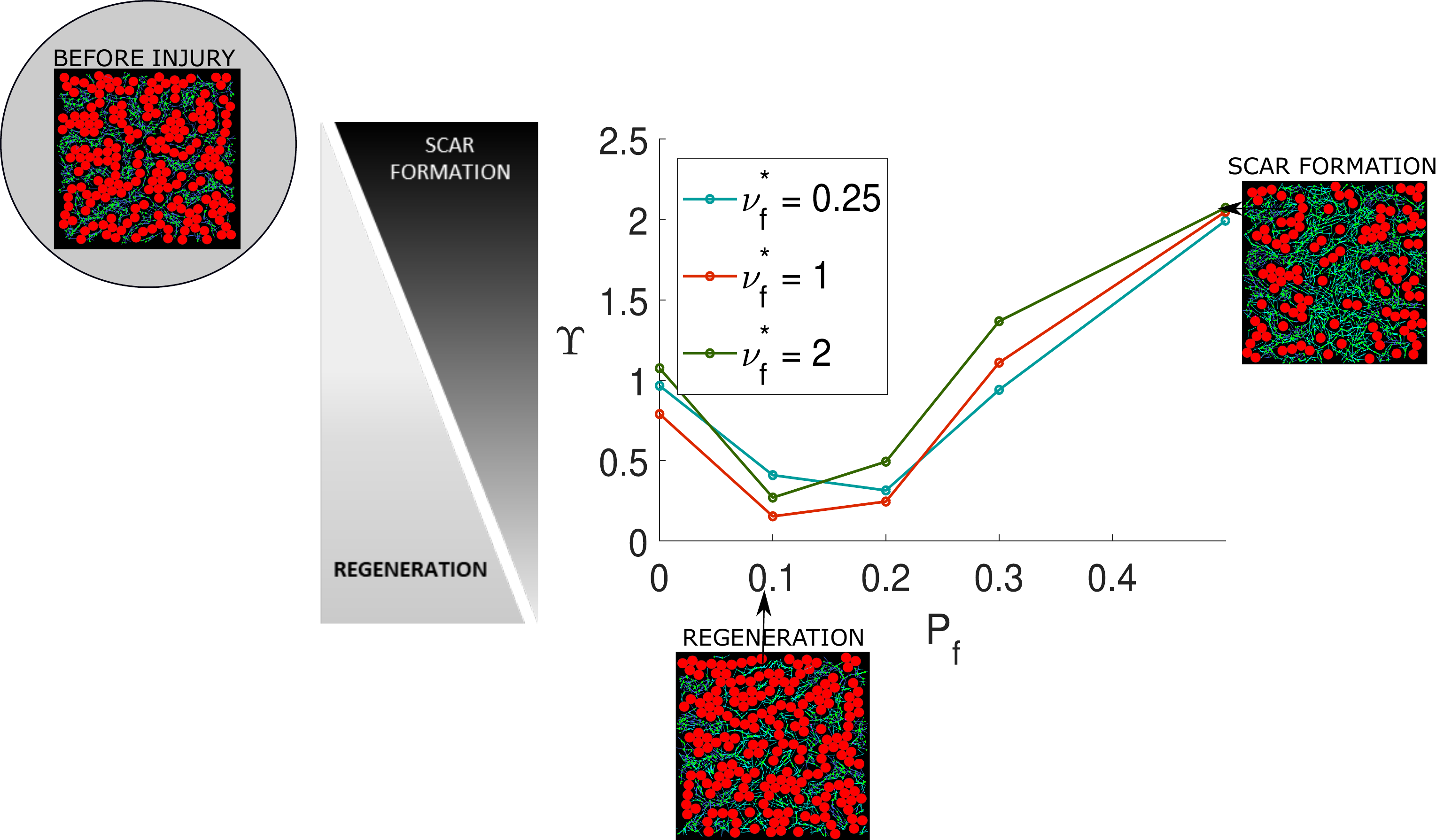}
\caption{Values of the global error $\Upsilon$ as a function of the linking probability of newly inseminated fibers $\mathbf{P}_f$, for different values of fiber insemination rates after injury $\nu_f^*$: $\nu_f^*=0.25$ (blue curve), $\nu_f^*=1$ (red curve) and $\nu_f^*=2$ (green curve), and for tissues with low remodelling rate $\chi_\ell=0.1, \nu_d = 10^{-3}$. \label{Fig:GEnud1e3}}
\end{figure}
In Fig. \ref{Fig:GEnud1e3}, we show the values of the global error $\Upsilon$ as a function of the linking probability of newly inseminated fibers $\mathbf{P}_f$, for three different fiber insemination frequencies after injury: $\nu_f^*=0.25$ (blue curve), $\nu_f^*=1$ (red curve) and $\nu_f^*=2$ (green curve), and for 'regular tissues' i.e with $\chi_\ell = 0.1$ and $\nu_d = 10^{-3}$.  Figs. \ref{Fig:GEnud1e2} and \ref{Fig:GEnud1e1} show the simulation results for $\chi_\ell = 0.35$ and $\nu_d = 10^{-2}$ (Fig. \ref{Fig:GEnud1e2}) and $\nu_d = 10^{-1}$ (Fig. \ref{Fig:GEnud1e1}). As shown in Fig. \ref{Fig:GEnud1e3}, the regeneration phase (where $\Upsilon$ is low) is observed for $\mathbf{P}_f \in [0.1, 0.2]$ and better regeneration is obtained for a moderate fiber insemination frequency $\nu_f^*=1$ compared to the cases $\nu_f^*=0.25$ or $\nu_f^* = 2$. Regeneration is therefore obtained with a moderate ECM reconstruction speed and when the newly inseminated fibers are sufficiently linked. For $\mathbf{P}_f=0$ indeed, the new tissue after injury is unstructured, showing that the new ECM must be sufficiently linked to enable tissue regeneration. However if the new ECM is too linked (large values of $\mathbf{P}_f$), we observe the formation of a structure like scar, i.e a fibrotic tissue composed of highly interconnected fiber structures with very few adipocytes. Indeed, for 'regular tissues' with low remodelling rate $\nu_d = 10^{-3}$, the initially highly interconnected fiber network fails to self-organize and prevent new cells to appear, generating a scar. Note that the fiber insemination rate after injury $\nu_f^*$ has little influence on the global error $\Upsilon$ for small values of linking probability of newly inseminated fibers $\mathbf{P}_f$, while for larger values of $\mathbf{P}_f$, increasing the fiber insemination speed favors scar formation. The dependency of the scar formation on the fiber insemination speed is enhanced when considering  more dynamical fiber networks (Fig. \ref{Fig:GEnud1e2}). Indeed for larger linking/unlinking frequency $\nu_d=10^{-2}$, the fiber network is more able to organize even when initially highly interconnected (for $\mathbf{P}_f=0.5$). When the fiber insemination speed is low, the fiber structures can align and enable the apparition of new cells, preventing scar formation in this case and generating an unstructured tissue. 
\begin{figure}[h!]
\includegraphics[scale=0.4]{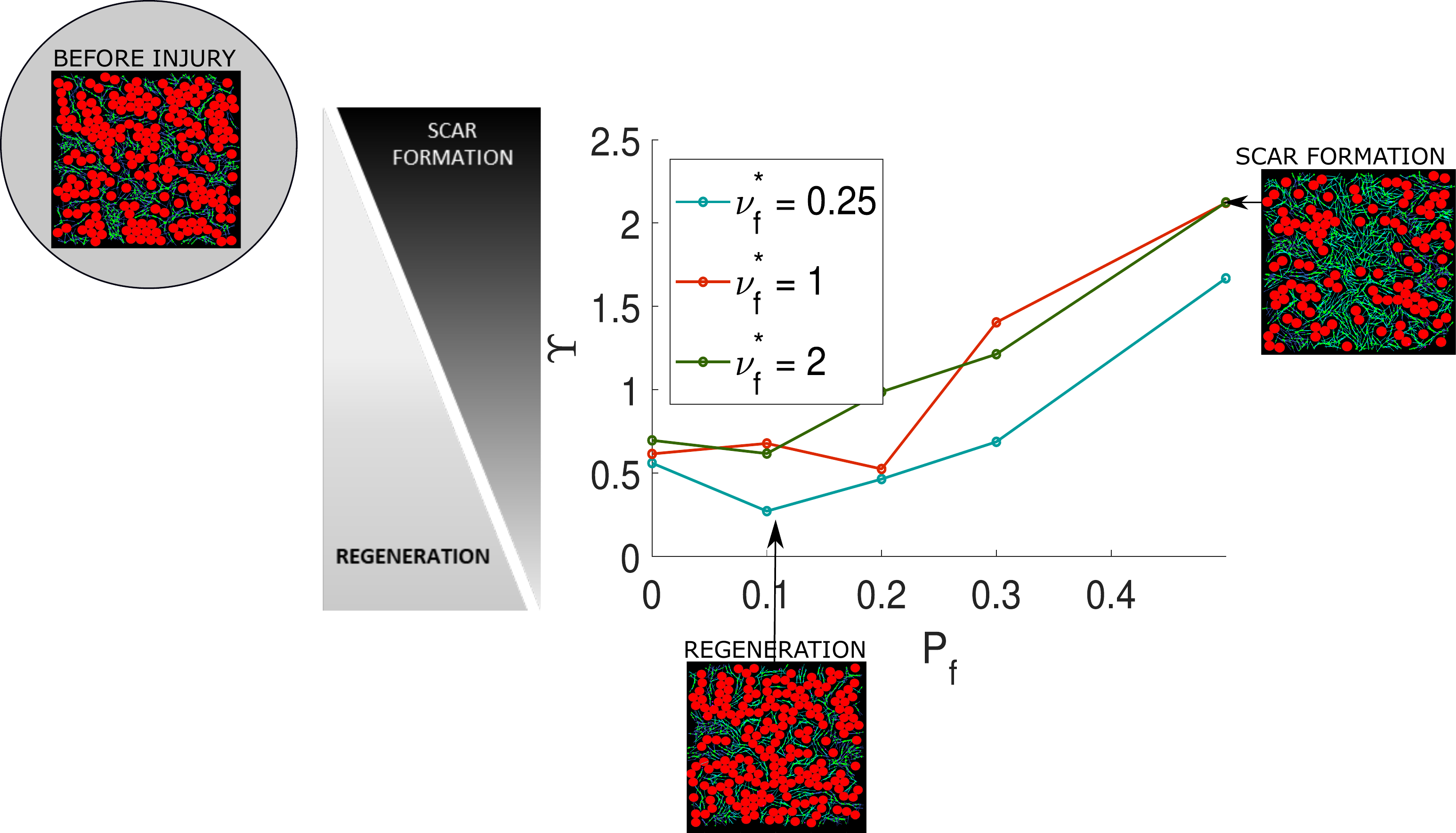}
\caption{Values of the global error $\Upsilon$ as function of the linking probability of newly inseminated fibers $\mathbf{P}_f$, for different values of fiber insemination rates after injury $\nu_f^*$: $\nu_f^*=0.25$ (blue curve), $\nu_f^*=1$ (red curve) and $\nu_f^*=2$ (green curve), and for tissues with moderate remodelling rate $\chi_\ell=0.35, \nu_d = 10^{-2}$. \label{Fig:GEnud1e2}}
\end{figure}
As shown in Fig. \ref{Fig:GEnud1e2} for moderately dynamical networks $\nu_d = 10^{-2}$, no regeneration is obtained for large fiber insemination speeds $\nu_f^* = 1$ or $2$, and regeneration is only obtained for $\nu_f^* = 0.25$ and $\mathbf{P}_f=0.1$. This shows that the reconstruction of more plastic fiber networks must be slower than for 'regular tissues' to enable regeneration to occur. 
\begin{figure}[h!]
\includegraphics[scale=0.4]{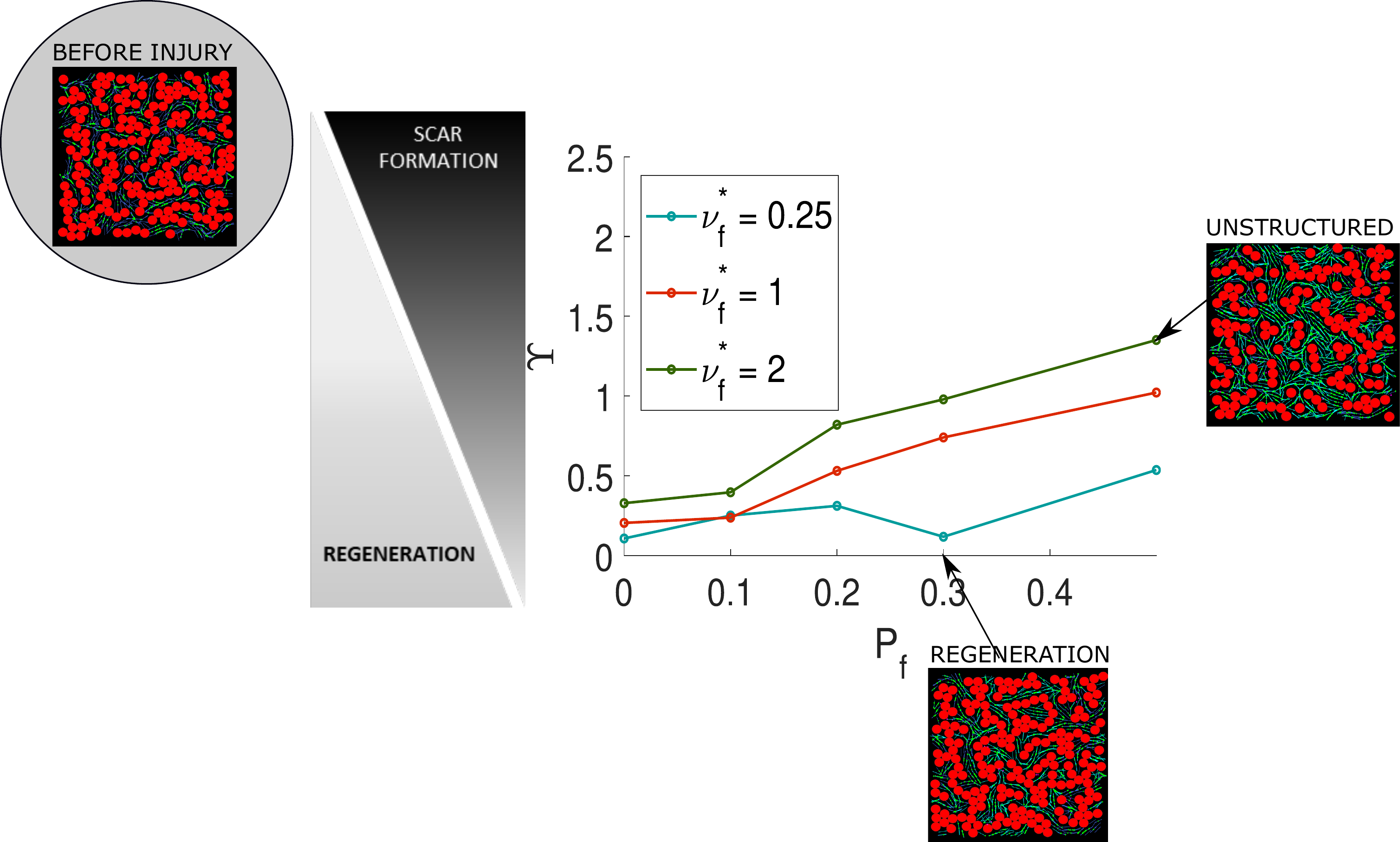}
\caption{Values of the global error $\Upsilon$ as a function of the linking probability of newly inseminated fibers $\mathbf{P}_f$, for different values of fiber insemination rates after injury $\nu_f^*$: $\nu_f^*=0.25$ (blue curve), $\nu_f^*=1$ (red curve) and $\nu_f^*=2$ (green curve), and for tissues with large remodelling rate $\chi_\ell=0.35, \nu_d = 10^{-1}$.\label{Fig:GEnud1e1}}
\end{figure}
When considering highly dynamical networks ($\nu_d = 10^{-1}$, Fig. \ref{Fig:GEnud1e1}), one can note that scar formation is not observed no matter the linking probability of newly inseminated fibers $\mathbf{P}_f$ nor the fiber insemination speed $\nu_f^*$. In this case indeed, as the fiber network remodelling is very fast, the fiber structures quickly self-organize and align, enabling the apparition of new cells. For slow fiber insemination, the state before injury is recovered independently of $\mathbf{P}_f$(regeneration), while larger fiber insemination speed lead to the formation of unstructured tissues. 

Finally, we show in Fig. \ref{Fig:GEnudstar1} the same computations as in Figs. \ref{Fig:GEnud1e3}-\ref{Fig:GEnud1e1} but with injury signaling particles lifetime increased by a factor 2 ($\nu_D^* = 1$ instead of $\nu_D^* = 0.5$ as before). As we can observe, we obtain the same results as before (compare Figs. \ref{Fig:GEnudstar1} (A) - (C) to Figs. \ref{Fig:GEnud1e3}, \ref{Fig:GEnud1e2}, \ref{Fig:GEnud1e1} respectively). These results show that in the regime of parameters chosen here, the lifetime of the injury signaling particles is not a key parameters in the fate of injury reconstruction. Note that the presence of injury signaling particles -blocking cell differentiation as long as they are present locally- together with fiber insemination are the only new model ingredients compared to the previous model for AT morphogenesis. The model results suggest that if these new ingredients are essential to enable tissue reconstruction after injury, they are not determinant in the choice of regeneration/scar formation. The characteristics of the reconstructed tissues seem to mostly be determined by the mechanical properties of the reconstructed ECM. 
\begin{figure}[h!]
\includegraphics[scale=0.4]{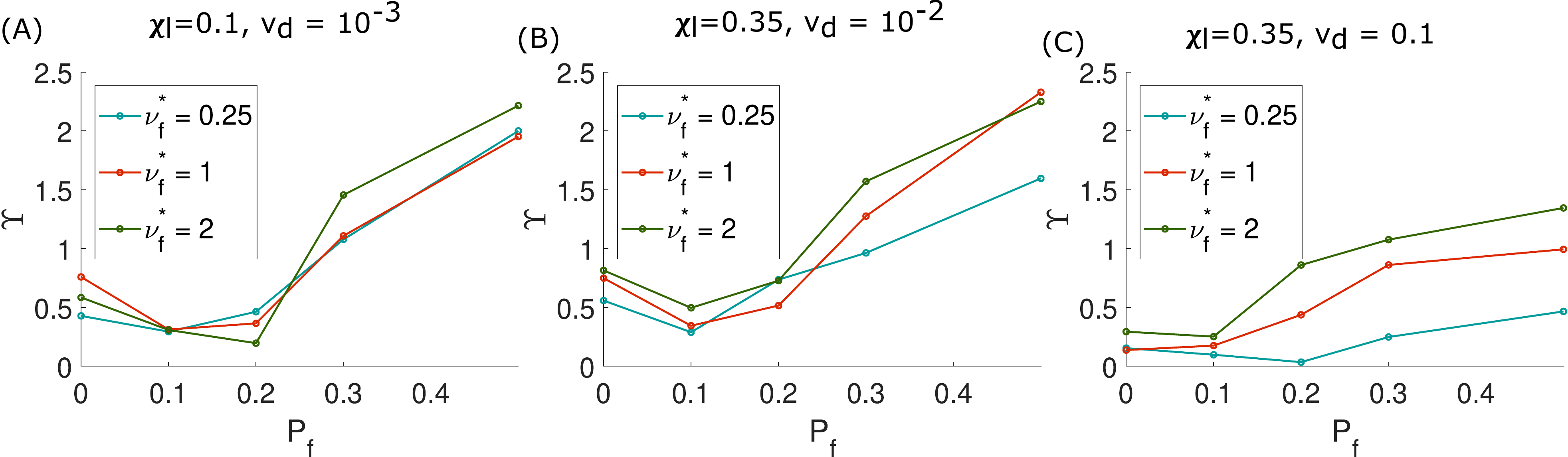}
\caption{Values of the global error $\Upsilon$ as a function of the linking probability of newly inseminated fibers $\mathbf{P}_f$, for different values of fiber insemination rates after injury $\nu_f^*$: $\nu_f^*=0.25$ (blue curve), $\nu_f^*=1$ (red curve) and $\nu_f^*=2$ (green curve), and for different tissues (low remodelling rate large $\chi_\ell=0.1, \nu_d = 10^{-3}$ (A), moderate remodelling rate $\chi_\ell=0.35, \nu_d = 10^{-2}$ (B), low remodelling rate $\chi_\ell=0.35, \nu_d = 10^{-1}$ (C)), and for injury signaling death rate $\nu_D* = 1$. \label{Fig:GEnudstar1}}
\end{figure}

\section{Discussion}

In our previous study [43], we demonstrated that AT morphogenesis results from a self-organization process principally driven by simple mechanical interactions between adipocytes and the ECM. Using the same framework, this study represents a step forward into the understanding of the mechanisms of AT repair after injury. The originality of our work lies in the hypothesis that the same mechanisms as for AT morphogenesis could be at play in tissue reconstruction after injury. Here, we showed that our mathematical model is either able to reproduce regenerated states or formation of scars after injury, by simply varying few key parameters. As a comparison, the main difference between the morphogenesis model and the tissue repair one is that tissue injury is represented as a gap in the tissue, whereas the tissue filled all the space in the morphogenesis model. This means that new components (cells and fibers) appear to fill the gap.

The states after injury that emerge from the mathematical model can be classified into three types: (i) Regeneration, i.e. complete reconstruction of the tissue after injury with the same number of adipocytes and formation of lobule-like clusters of cells in the new network, (ii) formation of an unstructured tissue with the same number of cells as before injury, but which do not display any organization of the cell and fiber clusters and (iii) scar formation, where the injured zone is filled with highly interconnected and disorganized fibers with very few adipocytes. Each state corresponds to a range of model parameters among which the rate of new fiber insemination $\nu_f^*$ and the linking probability between fibers at insemination $P_f$ seem to play a crucial role (Figure 1,2,3,4).  To weigh the relative roles of these two parameters in driving tissue regeneration or scar formation, we ran our model in three different tissues exhibiting low, moderate or high remodeling rates respectively (Figure 5,6,7). This remodeling rate correspond to the unlinking frequency $\nu_d$ for a fixed fraction of linked fibers $\chi_\ell$ (the larger the unlinking frequency the larger the linking frequency as $\nu_\ell = \frac{\chi_\ell}{1-\chi_\ell} \nu_d$). Regardless of the rate of new fiber insemination, the outcome of tissue repair is the same in a tissue exhibiting low fiber linking-unlinking dynamic (enabling low remodeling of fiber structures). In such tissues, the key parameter driving tissue reconstruction is the linking probability of new fibers $P_f$ as tissue regeneration is favored in a range of $P_f$ comprised between 0.1 and 0.2 whereas, higher values of $P_f$ are associated with scar formation. In tissues exhibiting moderate or high remodeling capabilities, a large rate of new fiber insemination decreases tissue regeneration capabilities. Moreover, under conditions of low rate of new fiber insemination, regeneration can be obtained for a wider range of linking probabilities of newly inseminated fibers ($P_f$ comprised between 0.1 and 0.3) compared to tissues exhibiting low remodeling rates. For higher fiber insemination rates, we observe either the formation of an unstructured tissue or of a scar. This observation is related to the fact that highly dynamical tissues exhibiting large plasticity quickly remodel their fiber network allowing the creation of organized (aligned) fiber structures as long as fiber density is not too high. All these tissue features provide ECM physical properties which control stem cell differentiation through cell-matrix protein interactions. Thus, in our model, a low fiber insemination rate associated to a high plasticity of the ECM induce the creation of new adipocytes in AT and facilitate tissue regeneration compared to tissues with lower remodeling rate. 

Our model suggests that there exists a value of ECM link proportion at creation after injury $P_f=0.1$ for which the tissues regenerate, independently of other parameters (i.e independently of the type of tissue -exhibiting low, moderate or high ECM remodelling abilities- or of the fiber insemination frequency $\nu_f^*$). On the other hand, the capability of a tissue to generate a scar is multifactorial. Indeed, scar formation was only observed for tissues with low or moderate remodelling abilities and for which the proportion of fiber links at creation is above a given threshold. This threshold itself depends on the fiber insemination rate and the type of tissue considered. Moreover, qualitatively, this study showed that the same model can be used to study tissue homeostasis or induce regeneration. On the other hand, quantitatively, our work showed that tissue regeneration is possible for specific values of the ECM link proportion at creation $P_f$ which is lower than the one corresponding to ECM of the homeostasis phase (before injury).

In conclusion, our results show that ECM reconstruction capabilities are key elements in the fate of injury outcome, thereby highlighting the importance of the ECM and the mechanical forces induced by it. They are consistent with recent papers demonstrating the impact of the ECM and of its dysregulation (i.e. fibrosis/scar) on repair processes \cite{Friedl_2000,Khan_2009,Pellegrinelli_2014}. 

Further improvements of the model could be made. From a mathematical viewpoint, the development of a macroscopic model from the present IBM would enable simulations on larger domains and address the question of the organization of the whole tissue. A first step has been made in \cite{Cristancho_2011} where a macroscopic model for interacting fibers has been derived. This macroscopic model has been further analyzed and its derivation has been numerically validated in \cite{Peurichard2016}. Moreover, the model could also be extended to 3D without profound methodological modifications (see discussion in \cite{Peurichard2016}). From a biological viewpoint, the development of a model including a time scale similar to what is observed in vivo in mouse models of AT regeneration and scar healing should allow us to define a precise time window during which pharmacological treatments would act on ECM fiber linking and in fine drive tissue repair towards regeneration. 

\section*{Acknowledgments}
PD acknowledges support from the Engineering and Physical Sciences Research Council (EPSRC) under grant ref. EP/M006883/1 and EP/N014529/1, from the Royal Society and the Wolfson foundation through a Royal Society Wolfson Research Merit Award no. WM130048 and from the National Science Foundation (NSF) under grant RNMS11-07444 (KI-Net). PD is on leave from CNRS, Institut de Math\'ematiques, Toulouse, France. D.P. acknowledges support by the Vienna Science and Technology Fund (WWTF) under project number
LS13-029, and gratefully acknowledges the Institut de Math\'ematiques de Toulouse (IMT) for sharing computing resources enabling the present study. 

\section*{Data statement} 
No new data were collected in the course of this research.

\appendix

\section{Supplementary material: videos of numerical simulations}
\subsection{S1 Video}
\label{S1_Vid}
{\bf Simulation: Reconstruction of 'regular tissues' ($\chi_\ell = 0.1, \nu_d = 10^{-3}$) after injury, with slow fiber reconstruction rate and moderate linking of new fibers}. 
This video shows a simulation of the reconstruction steps of a 'regular tissue' ($\chi_\ell = 0.1, \nu_d = 10^{-3}$), with slow fiber reconstruction rate and moderate linking of new fibers after injury: $\nu_f^* = 0.25$ and $\mathbf{P}_f=0.2$. Cells are represented as 2D spheres (red), the fibers before injury in dark blue and new fibers in cyan, fiber links are represented as green dots and injury signalling particles as magenta dots. At time $t=1$ the tissue at equilibrium is cut in half and we let the tissue reconstruct during 100$t_{ref}$ (400 days). Left panel: Initial condition (tissue at homeostasis before injury). Center panel: tissue reconstruction after injury (the injury zone is marked by white dashed lines). Right panel: evolution in time of the global error $\Upsilon$ which measures the distance between the tissue quantifiers at time $t$ and before injury (small values of $\Upsilon$ indicate a regeneration while large values indicate scar formation). At equilibrium, the tissue is reconstructed with structures similar to the ones before injury, showing that this set of parameters lead to regeneration for regular tissues.

\subsection{S2 Video}
\label{S2_Vid}
{\bf Simulation: Reconstruction of 'regular tissues' ($\chi_\ell = 0.1, \nu_d = 10^{-3}$) after injury, with large fiber reconstruction rate and large linking of new fibers}. 
This video shows a simulation of the reconstruction steps of a 'regular tissue' ($\chi_\ell = 0.1, \nu_d = 10^{-3}$), with large fiber reconstruction rate and large linking of new fibers after injury: $\nu_f^* = 2$ and $\mathbf{P}_f=0.5$. Cells are represented as 2D spheres (red), the fibers before injury in dark blue and new fibers in cyan, fiber links are represented as green dots and injury signalling particles as magenta dots. At time $t=1$ the tissue at equilibrium is cut in half and we let the tissue reconstruct during 100$t_{ref}$ (400 days). Left panel: Initial condition (tissue at homeostasis before injury). Center panel: tissue reconstruction after injury (the injury zone is marked by white dashed lines). Right panel: evolution in time of the global error $\Upsilon$ which measures the distance between the tissue quantifiers at time $t$ and before injury (small values of $\Upsilon$ indicate a regeneration while large values indicate scar formation). At equilibrium, the injury zone is filled with a highly connected fiber network possessing no adipocytes, which resemble a scar.

\subsection{S3 Video}
\label{S3_Vid}
{\bf Simulation: Reconstruction of 'highly dynamical tissues' ($\chi_\ell = 0.35, \nu_d = 10^{-1}$) after injury, with slow fiber reconstruction rate and moderate linking of new fibers}. 
This video shows a simulation of the reconstruction steps of a 'highly dynamical tissue' ($\chi_\ell = 0.35, \nu_d = 10^{-1}$), with slow fiber reconstruction rate and moderate linking of new fibers after injury: $\nu_f^* = 0.25$ and $\mathbf{P}_f=0.2$. Cells are represented as 2D spheres (red), the fibers before injury in dark blue and new fibers in cyan, fiber links are represented as green dots and injury signalling particles as magenta dots. At time $t=1$ the tissue at equilibrium is cut in half and we let the tissue reconstruct during 100$t_{ref}$ (400 days). Left panel: Initial condition (tissue at homeostasis before injury). Center panel: tissue reconstruction after injury (the injury zone is marked by white dashed lines). Right panel: evolution in time of the global error $\Upsilon$ which measures the distance between the tissue quantifiers at time $t$ and before injury (small values of $\Upsilon$ indicate a regeneration while large values indicate scar formation). At equilibrium, the tissue is reconstructed with structures similar to the ones before injury, showing that this set of parameters lead to regeneration for highly dynamical tissues.

\subsection{S4 Video}
\label{S4_Vid}
{\bf Simulation: Reconstruction of 'highly dynamical tissues' ($\chi_\ell = 0.35, \nu_d = 10^{-1}$) after injury, with large fiber reconstruction rate and large linking of new fibers}. 
This video shows a simulation of the reconstruction steps of a 'highly dynamical tissue' ($\chi_\ell = 0.35, \nu_d = 10^{-1}$), with large fiber reconstruction rate and large linking of new fibers after injury: $\nu_f^* = 2$ and $\mathbf{P}_f=0.5$. Cells are represented as 2D spheres (red), the fibers before injury in dark blue and new fibers in cyan, fiber links are represented as green dots and injury signalling particles as magenta dots. At time $t=1$ the tissue at equilibrium is cut in half and we let the tissue reconstruct during 100$t_{ref}$ (400 days). Left panel: Initial condition (tissue at homeostasis before injury). Center panel: tissue reconstruction after injury (the injury zone is marked by white dashed lines). Right panel: evolution in time of the global error $\Upsilon$ which measures the distance between the tissue quantifiers at time $t$ and before injury (small values of $\Upsilon$ indicate a regeneration while large values indicate scar formation). Contrary to 'regular tissues' (see Video \ref{S2_Vid}), scar formation is not observed for highly dynamical tissues. By rapidly creating and suppressing new links, ECM networks of highly dynamical tissues have a larger plasticity than the 'regular tissues', enabling new adipocytes to appear. 

\section{Immunofluorescence and biphoton microscopy.}\label{method} { Mouse inguinal adipose tissues (AT) were fixed, embedded in agarose gel and cut into $300 \mu m$ slices. Slices were incubated in Bodipy in PBS/$0.2\%$ triton 2h at room temperature. Imaging was performed using a Biphotonic Laser Scanning microscope (LSM880 – Carl Zeiss, Jena, Germany) with an objective lens LCI 'Plan Apochromat' 10x/0.45 and excited using 561 and 800 nm lasers. }

\section{Details of the components of the model}\label{Appendix:Model}
In this section, we detail the model ingredients for the simulations shown in the main text, and we refer to \cite{Peurichard2016} for more details. The two-dimensional Individual Based Model (IBM) consists of $N_A$ adipocytes, which are modelled as 2D growing spheres of center $X_i$ and radius $R_i$ for $i$ in $[1,N_A]$, and $N_f$ extra-cellular-matrix fiber elements which are represented by straight segments of fixed length $L_f$, of center $Y_k$ and orientation angle $\theta_k$ for $k \in [1,N_f]$. Adipocytes are prevented to overlap to model volume exclusion between cells. Fiber elements have the ability to link to or unlink from each other to model fiber elongation or rupture. The resistance of fibers to growing adipocytes is modelled by a repulsion potential $W_{pot}$ between cells and fibers, and fibers are supposed to repel each other by the same potential. Additionally, fibers offer resistance to bending through an alignment potential $W_{al}$ acting between two linked fiber elements. Cells and fibers seek to minimize their mechanical interaction energy resulting from these potentials, subject to the non-overlapping constraint between cells and to the linkage constraint between linked fibers. In the course of the simulation, an injury occurs, new cross-linked fibers are inseminated, new adipocytes appear as a result of stem-cell differentiation, adipocytes grow, new fiber links can appear and existing links can disappear. These phenomena disrupt the mechanical equilibrium and force the cells and fibers to move in order to restore the equilibrium.

\subsection{Mechanical potential}\label{App:Mechanics}
 Given a configuration at a fixed time, we denote by $C$ the set of cell center positions and radii: $C=\{(X_i, R_i) \; , \; i \in [1,N_A]\}$ and by $F$ the set of fiber center positions and fiber directional angles: $F = \{ (Y_k, \theta_k) \; , \; k \in [1,N_f]\}$. The global mechanical energy of the system reads: 
\begin{equation}
\displaystyle \mathcal{W}(C,F) = W_{\text{pot}} (C,F) + W_{\text{al}} (F) ,\label{pot}
\end{equation}
\noindent where $W_{\text{pot}}$ contains the fiber-cell and fiber-fiber repulsion potentials, and $W_{\text{al}}$ is the linked fiber-fiber alignment potential:
\begin{align}
& W_{pot} (C,F) = \sum_{1 \leq i \leq N_A} \sum_{1 \leq k \leq N_f} W (X_i,Y_k,\theta_k, d_{0,i}, \tilde{W}_k) &+ \sum_{1\leq f \leq N_f}\sum_{1\leq k \leq N_f} W(Y_k,Y_f,\theta_f, 2d_0,\tilde{W}) \nonumber\\
& &+ W(Y_f,Y_k,\theta_k, 2d_0, \tilde{W}) \label{potpot}\\
& W_{al}(F) =  c_1 \sum_{1\leq k \leq N_f}\sum_{1\leq m \leq N_f} p^t_{km} \sin^2(\theta_k - \theta_m).&\label{potal} 
\end{align}
\noindent The time-dependent coefficients $p^t_{km}$ are set to 1 if fibers $k$ and $m$ are linked at time $t$ and to 0 otherwise. The alignment potential $W_{al}$ of intensity $c_1$ consists of the sum of elementary alignment potentials between fibers of a linked pair. The repulsion potential $W_{pot}$ is supposed to be the sum of two-particle potential elements, $W(X_i,Y_k,\theta_k, d_{0,i}, \tilde{W}_k)$, modeling the mechanical interaction between cell $i$ and fiber $k$, and the sum of two-fiber potential elements $W(Y_f,Y_k,\theta_k, 2d_0, \tilde{W})$, modeling the mechanical interaction between fiber $k$ and fiber $f$. The isolines of these potential elements consist of ellipses with foci located at the two ends of the fiber segment corresponding to the second and third variables of the functional $W$. Given two vectors $X$, $Y \in $ $\mathbb{R}^2$, an angle $\theta\in [-\frac{\pi}{2},\frac{\pi}{2}]$, a distance $d_0$ and an intensity $\tilde{W}$, $W(X,Y,\theta,d_0,\tilde{W})$ reads:  
\begin{equation}\label{Wrep}
  W(X,Y,\theta,d_0,\tilde{W})=\begin{cases}  \frac{\tilde{W}}{g(d_0)} \big(g(d_0) - d(X,Y,\theta)\big) \; \text{ if $d(X,Y,\theta)\leq g(d_0)$}\\
  0 \hspace{3,2cm}\text{ otherwise}
  \end{cases}
\end{equation}
\noindent where:
\begin{equation*}
d (X,Y,\theta) = |X - Y + \frac{L_f}{2} \omega(\theta)|+|X - Y - \frac{L_f}{2} \omega(\theta)|,
\end{equation*}
\noindent and $\omega(\theta) = \begin{pmatrix} \cos \theta \\ \sin \theta \end{pmatrix}$ is the unit vector associated to the fiber angle $\theta$. We suppose that  fiber $k$ can repel cell $i$ and fiber $f$ up to a distance $d_0$ in its orthogonal direction. Note that fiber-cell repulsion (resp. fiber-fiber repulsion) aims to model repulsion of the fiber by the border of the cell (resp. by the border of the other fiber), while a cell (resp. a fiber) is modeled by its center. The repulsion distance is accordingly chosen to $d_{0,i} = d_0 + R_i$ for a cell and to $2d_0$ for a fiber (see Fig. \ref{fiberrep}). A direct computation gives: 
\begin{equation}\label{d0}
\begin{split}
g(d_0) = 2\sqrt{\frac{L_f^2}{4} + d_0^2}.
\end{split}
\end{equation}
\noindent Finally, we let the strength of each cell-fiber repulsion potential element $\tilde{W}_k$ depend on the local alignment of the fiber network \cite{Friedl_2000}. To this aim, we let $\tilde{W}_k$ be a function comprised between $W_0$ and $W_1$, equal to $W_0$ when the local alignment around fiber $k$ is weak and $W_1$ when it is strong (see \cite{Peurichard2016} for more details). 

\begin{figure}
\includegraphics[scale=0.45]{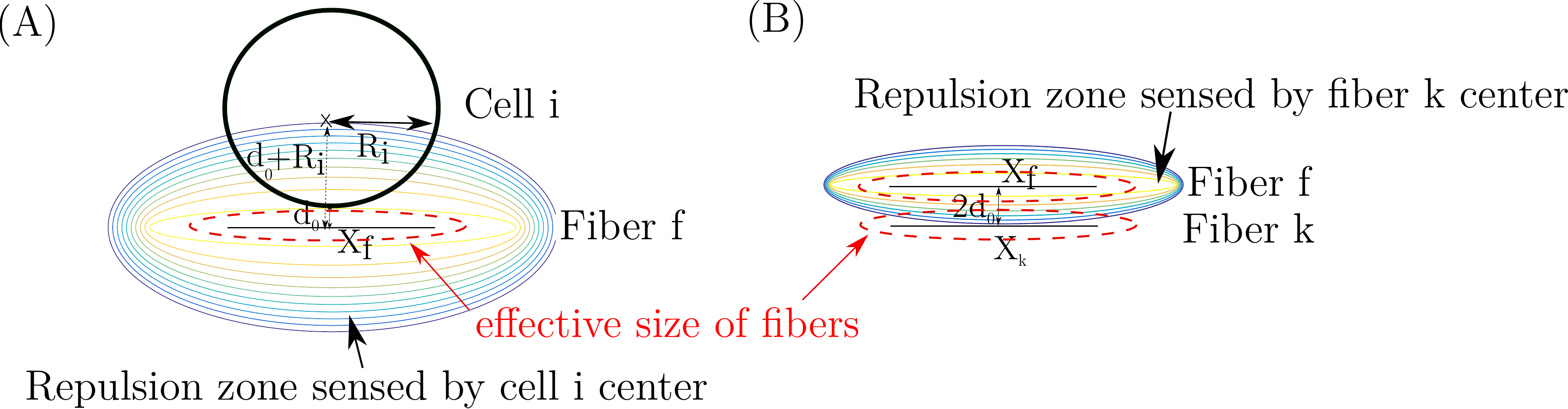}
\caption{Scheme of the repulsion potential exerted by fiber $f$ on (A) cell $i$, with orthogonal distance $d_0+R_i$ from fiber $f$ center to cell $i$ center and (B) fiber $k$, orthogonal distance $2d_0$ from fiber $f$ center to fiber $k$ center. The isolines of the fiber $f$ repulsion potential are indicated in color and the effective size of fibers (inside which no overlapping occurs with any other part of the neighboring agents) is indicated in red dashed lines.\label{fiberrep}}
\end{figure}
At each time step, the minimization of the global mechanical energy under the constraints reads:
\begin{equation}
\label{argmin}
(C,F) = \underset{\Phi(\tilde{C})\leq 0 , \; \Psi(\tilde{F})=0}{\textrm{argmin}} \mathcal{W}(\tilde{C},\tilde{F}),
\end{equation}
\noindent where $\Phi(\tilde{C})\leq 0$ expresses cell-cell non-overlapping inequality constraints (iii) and   $\Psi(\tilde{F})=0$ expresses fiber-fiber linkage equality constraints (v) (further detailed), where:
\begin{equation*}
\begin{split}
 \Phi(C) &= \big( \Phi_{ij}(X_i,X_j)\big)_{(i,j) \in [1,N_A]^2}, \\
 \Psi(F) &=  \big(\vec{\Psi}_{km}(Y_k,Y_m,\theta_k,\theta_m) \big)_{(k,m) \in \mathcal{N}_f}.
 \end{split}
 \end{equation*}
\noindent  Here, $\mathcal{N}_f$ denotes the dynamical set of linked fiber pairs: $\mathcal{N}_f = \{ (k,m) \in [1,N_f]^2 , \, k < m , \, p_{km}^t = 1\}$. 

\subsection{Constraints}\label{App:Constraints}
 The non overlapping constraint between cells $i$ and $j$ is written as an inequality constraint on the following function $\Phi_{ij}$: 
\begin{equation}\label{Phi}
\Phi_{ij}(X_i,X_j)=(R_i+R_j)^2 - |X_i-X_j|^2.
\end{equation}
\noindent One immediately notes that cells $i$ and $j$ do not overlap if and only if $\Phi_{ij}(X_i,X_j) \leq 0$. 

To model fiber growth and elongation or conversely rupture, unlinked (resp. linked) intersecting fibers have the possibility to link (resp. unlink) at random times. As long as a pair of linked fibers remains linked, the attachment sites of the two linked fibres are kept at the same point. The maitenance of the link between fibers $k$ and $m$ is modeled as an equality constraint $\vec{\Psi}_{km}=0$ with:
\begin{equation}\label{P}
\vec{\Psi}_{km} (Y_k,Y_m,\theta_k,\theta_m) = Y_k + \ell_{km} \omega(\theta_k) - Y_m - \ell_{mk} \omega(\theta_m),
\end{equation}
\noindent where $\ell_{km}$ is the distance of the center of fiber $k$ to its attachment site with fiber $m$ (see \cite{Peurichard2016} for more details). 

\subsection{Biological Phenomena}
In this section, we give details on the treatment of the biological phenomena of the model (cell and fiber insemination and insemination of injury signaling chemicals after injury). The insemination locations of new agents (cells and fibers) are supposed to depend on some local densities of existing agents at the time of their creation: fibers and injury signaling chemicals are created at the -moving- edges of the tissue, and cells are inseminated in an existing ECM which is not too linked locally. To this aim, we first describe how we compute the densities of existing agents at a given time of the simulation, and then describe the treatment of each biological phenomenon.

\textbf{Computation of the agent density}
The agent density (cells and fibers) around the geometrical point $X \in \Omega$ is computed thanks to a Particle-In-Cell (PIC) method. It indicates the number of agents per unit volume. To compute it, the computational domain $\Omega$ is first discretized into $N_x+1$ (resp $N_y+1$) nodes in the $x$-direction (resp. $y$-direction) regularly distributed with distance $\Delta x$ (resp. $\Delta y$) from their neighbors, with $N_\Omega = 4 N_x N_y$ number of numerical boxes for the PIC method. Then, the contribution of each agent $i$ (cell or fiber) located at $(x_i,y_i)$ is distributed among the nodes surrounding $(x_i,y_i)$ such as in Fig. \ref{schemePIC}.
\begin{figure}[h]
\includegraphics[scale= .3]{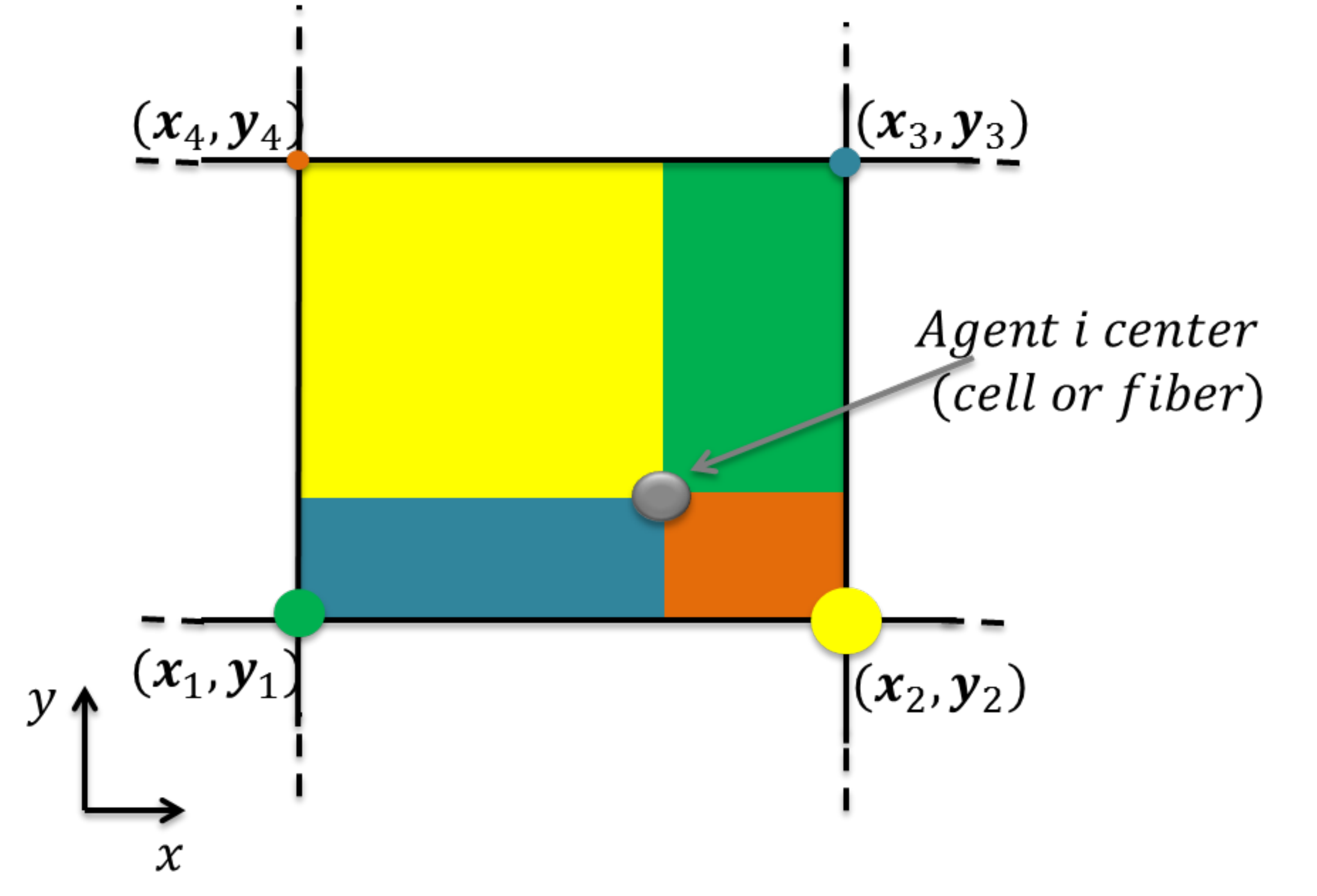}
\caption{Distribution of agent $i$ center (in gray) among the surrounding nodes $(\textbf{x}_1,\textbf{y}_1) ,(\textbf{x}_2,\textbf{y}_2),(\textbf{x}_3,\textbf{y}_3),(\textbf{x}_4,\textbf{y}_4)$. The weights of each node corresponds to the area with matching color. \label{schemePIC}}
\end{figure} 
Denoting $(\textbf{x}_1,\textbf{y}_1) ,(\textbf{x}_2,\textbf{y}_2),(\textbf{x}_3,\textbf{y}_3),(\textbf{x}_4,\textbf{y}_4)$ the four nodes (from the below left corner and numbered in trigonometric order) around $(x_i,y_i)$, $\rho(\textbf{x}_k,\textbf{y}_k)$ for $k = 1 \hdots 4$ is computed with:
\begin{equation*}
\rho(\textbf{x}_k,\textbf{y}_k) = \sum_{i | (x_i,y_i) \in [x_1,x_2]\times[y_1,y_2] } \frac{c_i}{(\Delta x \Delta y)^2} \omega_k(x_i,y_i),
\end{equation*}
\noindent where the weights $\omega_k$ for $k = 1 \hdots 4$ read:
\begin{align*}
\omega_1(x_i,y_i) = ( \textbf{x}_3 - x_i) ( \textbf{y}_3 - y_i)\\
\omega_2(x_i,y_i) = (x_i - \textbf{x}_4) ( \textbf{y}_4 - y_i)\\
\omega_3(x_i,y_i) = (x_i - \textbf{x}_1) (y_i - \textbf{y}_1)\\
\omega_4(x_i,y_i) = ( \textbf{x}_2 - x_i) (y_i - \textbf{y}_2).\\
\end{align*}
\noindent and $c_i$ is the agent-type dependent weight. As the maximal cell area is chosen to $\pi R_{\text{max}}^2\approx 1.4$ and the effective size of an individual fiber is set to $\pi \frac{L_f}{2} \frac{d_0}{2} \approx 0.36 \approx \frac{A_c}{5}$ (see Fig. \ref{fiberrep}), it is reasonable to consider that a cell occupies 5 times more space than an individual fiber. We therefore choose the weight $c_i = 1$ for a cell $i$, $c_i=\frac{1}{5}$ for a fiber $i$. The agent density biased by the local amount of linked fibers $\rho_\ell$ (used for defining the cell insemination frequency $\nu_a(X)$ of section 'Cell insemination') is computed in the same way, but letting the weight $c_i = \frac{N_{i,\ell}}{5}$ for a fiber $i$, where $N_{i,\ell}$ is the number of fibers linked to fiber $i$. Note that if fiber $i$ is not linked to any other fiber, $c_i=0$, and therefore the unlinked fibers do not contribute to the biased density of agents $\rho_\ell$. 

The gradient of $\rho$ is computed using finite differences on the numerical grid used for the PIC method (previously described). On the left boundary of the wound we use a forward second order finite difference scheme:
\begin{equation*}
\bigg(\frac{\partial \rho}{\partial x}\bigg)_i \approx \frac{-3\rho_{i} + 4 \rho_{i+1} - \rho_{i+2}}{2 \Delta x},
\end{equation*}
\noindent while for the right boundary we use a backward finite difference scheme:
\begin{equation*}
\bigg(\frac{\partial \rho}{\partial x}\bigg)_i \approx \frac{3\rho_{i} - 4 \rho_{i-1} + \rho_{i-2}}{2 \Delta x},
\end{equation*}
\noindent and finally in the center we use centered finite differences:
\begin{equation*}
\bigg(\frac{\partial \rho}{\partial x}\bigg)_i \approx \frac{\rho_{i+1} - \rho_{i-1}}{2 \Delta x}.
\end{equation*}

\textit{\bf Fiber insemination}
We suppose that fiber insemination follows a Poisson process in time, of spatially dependent frequency $\nu_f(X)  \in [0,\nu_f^*]$, $X \in \Omega$. In order to let the fiber insemination frequency depend on the local agent (cells of fibers) density $\rho$ as well as on the local agent density gradient length $g(\rho,\nabla \rho)$, inspired from \cite{Aymard2016}, we set:
\begin{equation*}
\nu_f(X) = \nu_f^* \psi\bigg(\frac{L_c^0 g\big(\rho(X),\nabla \rho(X)\big) - 1}{h_s}\bigg)\psi\bigg(\frac{1 - \frac{\rho(X)}{\rho_s}}{h_s}\bigg),
\end{equation*}
\noindent where $\nu_f^*,L_c^0, h_s,\rho_s$ are real positive model parameters, $\rho(X)$ is the agents density around point $X$ (further described) and
\begin{align*}
\psi(z) &= \frac{1}{2}(1+\tanh(z)) \text{  (smoothing function)}\\
g(\rho,\nabla \rho) &= \frac{||\nabla \rho ||}{\rho + \rho_0} \text{ \hspace{1cm} (inverse of gradient length)}.
\end{align*}
\noindent Here, $\rho_0$ is introduced to avoid singularities of the function $g(\rho,\nabla \rho)$ in cases where the local agents density $\rho$ is close to 0. The increasing smoothing function $\psi(\frac{z}{h_s}) \in [0,1]$ allows to control the dependency of the frequencies of the Poisson processes on the quantities contained in the variable $z$, with scaling control $h_s$ (see Fig. \ref{psi}). Thus defined, fiber insemination will occur more likely in zones of high agent gradient (close to the boundaries of the wound) and with low local agent density. The Poisson process is discretized as follows: we randomly pick up $N_\Omega$ points in the computational domain $\Omega$ and, for each point, we compute the value $\nu_f(X)$ and inseminate a fiber the center of which is at $X$ with probability $1-\exp^{-\nu_f(X) S \Delta t}$, where $S = \frac{|\Omega|}{N_\Omega}$. Note that $N_\Omega$ is the number of numerical boxes used to compute the agent density with the PIC method (explained further).

 \begin{figure}[h!]
\includegraphics[scale=0.2]{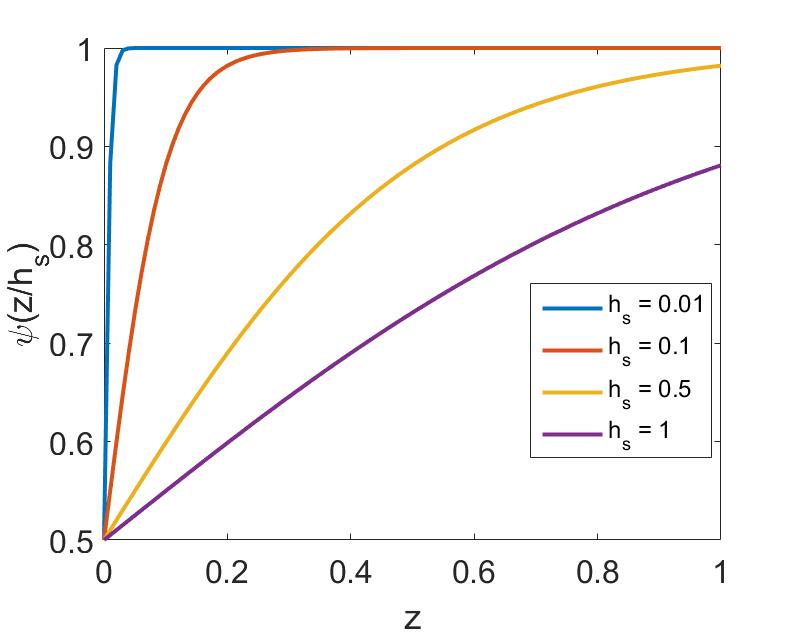}
\caption{Function $\psi(\frac{z}{h_s})$ as a function of $z\in [0,1]$, for different values of $h_s$: $h_s = 0.01$ (blue curve), $h_s = 0.1$ (orange curve), $h_s = 0.5$ (yellow curve) and $h_s = 1$ (purple curve).\label{psi}}
\end{figure}

\textit{\bf Injury signaling chemicals}
As for the fibers, we suppose that a flux of injury signaling chemicals (of density $\rho_R$) is created at the boundaries of the wound. We therefore inseminate these injury signaling chemicals following a Poisson process in time, of spatially dependent density $\nu_R(X)$, $X \in \Omega$:
\begin{equation*}
\nu_R(X) = \nu_R^* \psi\bigg(\frac{L_c^0 g\big(\rho(X),\nabla \rho(X)\big) - 1}{h_s}\bigg)\psi\bigg(\frac{1 - \frac{\rho(X)}{\rho_s}}{h_s}\bigg),
\end{equation*}
\noindent where $\nu_R^*>0$ is a parameter of the model. The Poisson process of frequency $\nu_R(X)$ is discretized as for the fiber insemination (see previous paragraph). We suppose that injury signaling particles have a lifetime $\frac{1}{\nu_D}$. Hence, the probability of suppressing an injury signalling particle between time steps $t$ and $t+\Delta t$ is $1-e^{-\nu_D\Delta t}$. Once created, these injury signaling particles are supposed to diffuse in the fibrous network in a directed manner, i.e the flux speed at location $X$ is supposed to be biased by the mean orientation of fibers around point $X$. The time evolution of the density of injury signalling particles then reads:
\begin{equation}\label{diffCont}
\partial_t \rho_R =  \nabla.\big(\rho_R D(X) \frac{\nabla \rho_R}{\rho_R + \rho_R^*} \big)
\end{equation}
\noindent where $\rho_R^*>0$ is a parameter introduced to avoid singularities where $\rho_R = 0$ and $D$ is a $2 \times 2$ diffusion matrix of the tissue around point $X$. In order to let injury signalling particles diffuse locally in a directed manner throughout the surrounding tissue, we let the diffusion matrix be the sum of the projection matrices on the direction vectors of the fibers contained in the ball centered in $X$ and of radius $\frac{L_f}{4}$:
$$
D(X) = d_0 \sum_{f \; | \; ||Y_f - X|| \leq \frac{L_f}{4}} \omega_f \otimes \omega_f,
$$
\noindent where $d_0>0$ is the diffusion parameter and $\omega_f = \begin{pmatrix} \cos \theta_f \\ \sin \theta_f \end{pmatrix}$ is the directional vector of fiber $f$. We consider Neumann boundary conditions on the right and left boundaries and periodic boundary conditions on the top and bottom. For any $y \in [-y_{\max},y_{\max}]$:
\begin{equation*}
\frac{\partial \rho_R}{\partial x}(x=-x_{\max},y) =  0, \qquad
\frac{\partial \rho_R}{\partial x}(x=x_{\max},y) =  0.
\end{equation*}
\noindent In order to compute the fluxes, we use a Smooth Particle Hydrodynamic approach \cite{Monaghan1988}, enabling the modelling of fluid elements as discrete objects. To this aim, let us denote by $Z$ the coordinate vector of an injury signalling particle. The Lagrangian description of the fluid following Eq. \eqref{diffCont} then reads:
\begin{equation}\label{dZdt}
\frac{d Z}{dt} = -D(X) \frac{\nabla\rho_R}{\rho_R + \rho_R^*}.
\end{equation}
\noindent Given a function $f(X)$, $X \in \Omega$ we have:
$$
f(X) = \int_\Omega f(X') \delta(X-X') dX'.
$$
\noindent At the discrete level, replacing the delta function by a smoothing kernel $W(X,h)$ with characteristic width $h$ such that:
$$
\underset{h\rightarrow 0}{\lim} W(X,h) = \delta(X),
$$
\noindent subject to the normalization
$$
\int_\Omega W(X',h) dX' = 1,
$$
\noindent we have:
$$
f(X) = \int_\Omega \frac{f(X')}{\rho_R(X')} W(X-X',h)\rho_R(X') dX' + O(h^2).
$$
\noindent After discretizing the continuous field in a series of $N$ particles $Z_j$ of mass $m_j$, $f(X)$ can be written:
$$
f(X) \approx \sum_{j=1}^N \frac{m_j}{\rho_j} f(Z_j) W(X-Z_j,h),
$$
\noindent and where $\rho_j \approx \rho_R(Z_j)$ and $\frac{m_j}{\rho_j}$ is the volume of a discrete particle $j$. The interpolated gradient of function $f$ then reads:
$$
\nabla f(X) = \sum_{j=1}^N \frac{m_j}{\rho_j} f(Z_j) \nabla W(X-Z_j,h).
$$
\noindent In our model, we use the poly6 kernel defined in 2D by:
$$
W(X,h) = \begin{cases} \frac{4}{\pi h^8} ( h^2 - ||X||^2)^3 \qquad \text{if } 0\leq ||X|| \leq h\\
0 \hspace{3cm} \text{otherwise},
\end{cases}
$$ 
\noindent the gradient of which reads:
$$
\nabla W(X,h) = \begin{cases} -\frac{24}{\pi h^8} ( h^2 - ||X||^2)^2 X \qquad \text{if } 0\leq ||X|| \leq h\\
0 \hspace{3cm} \text{otherwise}. 
\end{cases}
$$ 
\noindent The support radius $h$ is chosen such that the region of influence of each particle contains 10 to 20 neighbouring particles. Equation \eqref{dZdt} for each particle $(Z_j,m_j)\in \Omega \times \mathbb{R},j\in[1,N]$ then reads:
$$
\frac{dZ_j}{dt} = - D_j \frac{\nabla \rho_R(Z_j)}{\rho_j + \rho_R^*},
$$
 \noindent where $D_j = D(Z_j)$ and
 $$
 \nabla \rho_R(Z_j) = \sum_{k=1}^N m_k \nabla W(Z_j-Z_k,h).
 $$
We consider an explicit time discretization: for $t^{n} = \Delta t^1 + \hdots + \Delta t^n $, we write for every $j \in [1,N]$:
$$
Z_j^{n+1} = Z_j^n - \Delta t^n D_j^n \frac{\nabla \rho_R(Z^n_j)}{\rho^n_j + \rho_R^*},
$$
 where $\Delta t^n $ is time-dependent and respects the CFL condition:
$$
\Delta t^n \leq C \frac{h}{ \underset{j \in [1,N]}{\text{sup}} ||D_j^n \frac{\nabla \rho_R(Z^n_j)}{\rho^n_j + \rho_R^*}||},
$$
\noindent with $C \leq \frac{1}{2}$.

\textit{\bf Cell insemination}
Cell insemination is supposed to follow a Poisson process in time of space dependent frequency $\nu_a(X)$. In order to let cell insemination occur at locations where there is a low density of injury signaling particles $\rho_R$, enough agents  (cells or fibers) $\rho$ but not too much cells and linked fibers $\rho_\ell$, we define:
$$
\nu_a(X) = \nu_a^* \psi \big( \frac{1-2\frac{\rho_\ell(X)}{\rho_s}}{h_s}\big)\psi \big(\frac{1-2\frac{\rho_R(X)}{\rho_s}}{h_s}\big)\psi \big(\frac{2\frac{\rho(X)}{\rho_s}-1}{h_s}\big)
$$
\noindent where $\nu_a^*>0$ is a model parameter, and $\rho_\ell(X)$ is the density of agents biased by the local amount of linked fibers, computed as explained in the next section. New cells are inseminated with a small radius $R_i=0.1$.

\textit{\bf Cell growth}
After insemination, the volumes of the cells are supposed to grow linearly with time. Given a cell $i$ at time $t$, the radius of cell $i$ at time $t+ \Delta t$ reads:
\begin{equation*}
R_i^3(t+ \Delta t) = R_i^3(t) + K_{g} (1 + \eta \rho_{g}) 
\end{equation*}
\noindent where $\eta$ is a random number chosen uniformly in $[0,1]$ and $K_{g}$, $ \rho_{g}$ are two parameters such that $\frac{K_g}{\Delta t}$ is the mean volumetric cell growth per unit of time and $\frac{K_g \rho_g}{\Delta t}$ is related to the standard deviation of the volumetric cell growth per unit of time. The characteristic time of cell growth $t_{g}$ is defined as the mean time needed for a cell to reach its maximal radius $R_{\max}$ and reads:
\begin{equation*}
t_{g} = \frac{R_{\max}^3 \Delta t}{K_{g}}.
\end{equation*} 

\subsection{Statistical quantifiers}\label{computationSQ}
This section is devoted to the computation of statistical quantifiers used to describe cell and fiber structures in both numerical simulations and biological images. A cell cluster is defined as a set of cells almost in contact. Let $\sim_a$ be the reflexive and symmetric relation: 
\begin{equation*}
j \sim_a i \; \Leftrightarrow j \in \mathcal{N}_i,
\end{equation*}
\noindent where $\mathcal{N}_i$ is the set of cell $i$ neighbors:
\begin{equation}\label{ni}
\mathcal{N}_i = \{j \in [1,N_{a}] \; , j \neq i ,  \; | \; |X_i-X_j| \leq (R_i + R_j + \epsilon_a)^2\},
\end{equation}
\noindent where $\epsilon_a$ is the maximal allowed distance up to which two cells not in contact are defined as neighbors and is set to $50\% \max(R_i,R_j)$. The equivalence relation $\sim_A$ then reads:
\begin{equation*}
\begin{split}
j \sim_A i \Leftrightarrow & \exists n \in \mathbb{N}^*, \exists (a_1..a_n)\\
& \; \text{such that } j \sim_a a_1 \sim_a ... \sim_a a_n \sim_a i.
\end{split}
\end{equation*}
\noindent  Cells $i$ and $j$ belong to the same cluster if and only if $i \sim_A j$. 

The statistical quantifier $N_C$ is defined as the total number of cell clusters which have more than 5 adipocytes per 100 adipocytes.

The statistical quantifier $E$ measures the mean elongation of the cell clusters, and is defined as the number of cells at the boundary of the clusters  normalized by the total number of cells in the clusters. As the parameter $E$ is irrelevant for clusters with less than 5 cells, its computation is restricted for clusters $c$ such that $n_c >5$ and reads:
\begin{equation*}
E = \frac{\sum_{c=1}^{N_C} \textrm{Card}( \mathcal{R} \cap \mathcal{C}_c )}{\sum_{c=1}^{N_C} n_c}.
\end{equation*}
\noindent Here, $\mathcal{C}_c$ is the set of indices of the cells belonging to cluster $c$, $n_c$ is the number of cells in cluster $c$ and $\mathcal{R}$ is the set of indices of all cells with less than 5 neighbors: 
\begin{equation*}
\mathcal{R} = \{ i \in [1,N_A],\  \textrm{Card}(\mathcal{N}_i) \leq 5 \},
\end{equation*}
\noindent where $\mathcal{N}_i$ is defined by Eq. \eqref{ni}.

In order to describe the fiber structures, we define a fiber cluster as a set of neighboring quasi-aligned fiber elements. For this purpose, let us define $\mathcal{M}_k$ as the set of neighbors of fiber $k$, quasi-aligned with fiber $k$. Then:
\begin{multline*}
\mathcal{M}_k= \{m \in [1,N_{f}] \; , m \neq k ,\\  
 \min(d_{f}(Y_k, Y_m),d_{f}(Y_m, Y_k)) \leq 0 \\
  \text{ and } |\sin(\theta_k - \theta_m)|  < \sin(\frac{\pi}{4}) \},
\end{multline*}
\noindent where $d_f(Y_k, Y_m)$ reads:
\begin{equation*}
d_f(Y_k,Y_m) = d(Y_k,Y^-_m) + d(Y_k,Y^+_m) - 2 \sqrt{(\frac{L_f}{2})^2 + (\tau_f L_f)^2}.
\end{equation*}
\noindent Here, $ Y^\pm_m = Y_m \pm \frac{L_f}{2}\omega_m$ and $d(X,Y)$ is the distance of point $X$ to point $Y$. Note that $d_f(Y_k,Y_m) \leq 0$ (resp. $d_f(Y_m,Y_k) \leq 0$) if the center of fiber $k$ (resp. $m$) is contained in the ellipse of foci $Y^\pm_m$ (resp. $Y^\pm_k$) with semi minor axis of length $\tau_f L_f$ and semi major axis of length $\sqrt{(\frac{L_f}{2})^2 + (\tau_f L_f)^2}$. We chose $\tau_f = \frac{1}{3}$, which means that a fiber detects a neighboring fiber up to a distance $\frac{L_f}{3}$ in its orthogonal direction. This allows us to define fiber clusters as sets of quasi-aligned neighboring fibers. Let us define the reflexive and symmetric relation $\sim_f$ by: 
\begin{equation*}
k \sim_f m \; \Leftrightarrow m \in \mathcal{M}_k.
\end{equation*}
\noindent Define the equivalence relation $\sim_F$ such that:
\begin{equation*}
\begin{split}
k \sim_F m \Leftrightarrow & \exists n \in \mathbb{N}^*, \exists (a_1..a_n)\\
& \; \text{such that } k \sim_f a_1 \sim_f ... \sim_f a_n \sim_f m.
\end{split}
\end{equation*}
\noindent Then, we say that fibers $k$ and $m$ belong to the same cluster if and only if $m\sim_F k$. 

We define the SQ $A$ to quantify the mean alignment of the fibers of a cluster. Given a fiber cluster $c_f$, the mean alignment of its fibers is defined as the maximal eigenvalue $\lambda^+_{c_f}$ of the mean projection matrix on the direction vectors all fibers of cluster $c_f$:
\begin{equation*}
P_f = \frac{1}{n_f}\sum_{m \in c_f} \omega_{m} \otimes \omega_{m},
\end{equation*}
\noindent where $n_f$ denotes the number of fibers contained in the cluster $c_f$, and $\omega_{m}$ is the directional vector of fiber $m$. Then, $A$ is defined as the mean of the fiber cluster alignment, weighted by the number of fibers in the cluster:
\begin{equation*}
A = \frac{1}{N_F}\underset{1\leq c_f\leq N_{T_f}}{\sum} \lambda^{+}_{c_f} n_{c_f},
\end{equation*}
\noindent where $n_{c_f}$ is the number of fibers in cluster $c_f$.

\subsection{Parameters of the model}

If not otherwise stated, the model parameters used in the simulations are given by table \ref{Table}
\begin{table}[H]
\begin{tabular}{cccc}
Parameters & Numerical Value & Biological value & Description\\
\hline
 \rowcolor[RGB]{135,206,250}\multicolumn{4}{c}{Model parameters}\\
\hline
 \rowcolor[RGB]{210,210,210}\multicolumn{4}{c}{Agents}\\
\hline
$N_a$ & 190& N/A & Number of cells before injury\\
$R_{\text{max}}$ & 0.66 & 30 $\mu m$ & Maximal cell radius\\
$N_f$ & $1600$ & N/A &Number of fibers before injury\\
$L_f$ & 1.2 & 60 $\mu m$ &Fiber length\\
\hline
 \rowcolor[RGB]{210,210,210}\multicolumn{4}{c}{Mechanical Interactions}\\
\hline
$d_0$ & 0.4 & 20 $\mu m$&  Fiber width\\
$W_0$ & $10$ &N/A & Minimal cell-fiber repulsion force\\
$W_1$ & $15$ &N/A& Maximal cell-fiber repulsion force\\
$\tilde{W}$ & $5$&N/A & Fiber-fiber repulsion force\\
\hline
 \rowcolor[RGB]{210,210,210}\multicolumn{4}{c}{Biological phenomena}\\
\hline
$\nu_f^*$ & adapted & between & Fiber insemination frequency\\
$\textbf{P}_f$ & adapted & N/A & New fiber linking probability\\
$\nu_R^*$ & 20 & 4 $days^{-1}$& Injury signalling particle insemination frequency\\
$\nu_D^*$ & 0.5 &0.1 $days^{-1}$& Injury signalling particle suppression frequency\\
$\nu_a^*$ & 0.2 & 0.04 $days^{-1}$& Cell insemination frequency\\
\hline
 \rowcolor[RGB]{135,206,250}\multicolumn{4}{c}{Numerical parameters}\\
\hline
$|\Omega|$ & 1300 &2.6 $mm^2$& Area of the square domain\\
$N_\Omega$ & 1444 & N/A & Number of numerical boxes for the PIC method\\
$h$ & $L_f/3$ &N/A& Radius for the SPH method\\
$L_c^0$ & 0.6 &N/A& Critical gradient length\\
$\rho_s$ & .5 &N/A& Critical agent density\\
$h_s$ & .1 &N/A& Length of the distribution $\psi$\\
\end{tabular}
\caption{Numerical and model parameters of the simulations of the main text. \label{Table}}
\end{table}
\bibliographystyle{ieeetr}

\end{document}